# Response Regimes in Equivalent Mechanical Model of Moderately Nonlinear Liquid Sloshing

*M. Farid[*] and O. V. Gendelman*

*Faculty of Mechanical Engineering, Technion – Israel Institute of Technology*

**contacting author*, maorfarid@gmail.com.

The paper considers non-stationary responses in reduced-order model (ROM) of partially liquid-filled tank under external forcing. The model involves one common degree of freedom for the tank and the non-sloshing portion of the liquid, and the other one – for the sloshing portion of the liquid. The coupling between these degrees of freedom is nonlinear, with the lowest-order potential dictated by symmetry considerations. Since the mass of the sloshing liquid in realistic conditions does not exceed 10% of the total mass of the system, the reduced-order model turns to be formally equivalent to well-studied oscillatory systems with nonlinear energy sinks (NES). Exploiting this analogy, and applying the methodology known from the studies of the systems with NES, we predict a multitude of possible nonstationary responses in the considered ROM. These responses conform, at least on the qualitative level, to the responses observed in experimental sloshing settings, multi-modal theoretical models and full-scale numeric simulations.

## 1. Introduction

Partially filled liquid storage tanks of different shapes are used in many fields of engineering , including vehicles, sea crafts, aircrafts, for the storage of various, maybe hazardous, liquids. The term “sloshing” refers to oscillatory relative motion of the liquid with respect to the containing vessel. The liquid sloshing may be hazardous for the vessel safety, since dynamic loads related to the sloshing may have direct and rather strong calamitous effect on the vessel stability and robustness.

While being most interesting and potentially hazardous, high-amplitude liquid sloshing in partially filled vessels still lacks thorough analytic representation. The sloshing liquid has infinite number of degrees of freedom; boundary conditions on the free surface are nonlinear and time-dependent. Near a resonance,one can encounter high-amplitude steep sloshing waves and hydraulic jumps. In the former case, nonlinear dynamical features may take place, for example multiple periodic solutions ('jump' phenomenon) [1], weakly quasi-periodic [2], [3] and strongly modulated responses [4]. Based on experimental results, it was pointed that cubic spring seems to describe the dynamical regimes in the best way [5]. When the hydraulic jumps are involved, significant hydraulic impacts are applied to the vessel inner walls [6]. Hydraulic jumps and wave collisions with the vessel shells express the strong non-linearities in the system [7]–[10].

Dynamics of the liquid sloshing in partially filled vessels depends on the liquid depth. For shallow free-surface, waves and hydraulic jumps show in vicinity of resonance and apply high loads on vessel walls. Lepelletier and Raichlen [11] show that in rectangular tanks under horizontal excitation experimental results are in good agreement with thelinear sloshing theory. Forced sloshing and nonlinear resonance phenomena were studied by Moiseev [12]. However, as the excitation frequency approaches the natural frequency of the lowest sloshing mode, nonlinear and dissipative theory must be used. Hydraulic jumps and wave collisions with vessel shells are the source of strong nonlinearities in the system. Moreover, they can cause significant local stresses in the containing structure and can affect the structure stability, resistance and robustness.

Due to computational complexity and cumulative error originated in the complex interaction between different bulks, the numerical solvers have limited ability to accurately describe the loads caused by the fluid-structure interactions over long time intervals. That is the motivation to look for analytical or semi-analytical methods. Comprehensive studies were made to identify and predict nonlinear sloshing regimes that appear near the resonance, with the help of asymptotic approaches. Most of those methods were based on assumptions of smooth container inner walls, and neglected viscosity [7], [8], [11]. Therefore, potential flow theory can be applied. In this work we focus on tanks with walls perpendicular to the undisturbed liquid free surface. Even though steady-state asymptotic modal theory might have a limited ability to quantitatively describe sloshing regimes in certain conditions (for example, when water impact on the tank top takes place), it may be instrumental for analysis of the moderately nonlinear responses, their stability and sensitivity to external perturbations. Moreover, the major advantage of the reduced-order model (ROM) is very small computational time compared to the CFD methods. Besides, the coupling between sloshing and the containing structure due to stochastic external excitation can be analyzed for long time intervals, and stability of the response regimes can be analyzed analytically.

Hill [13]and Hill and Frandsen [14] studied the wave height and nonlinear beating amplitude by solving analytically the third-order amplitude evolution equation is in terms of Jacoby elliptic functions and considering a single sloshing mode. Their solution is valid for finite depth and non-breaking waves. Steady-state asymptotic modal theory succeeded to predict and describe the observed and documented modulated, or beating sloshing phenomenon [2]. Stolbestov [15] theoretically studied the nonlinear sloshing regimes in almost-square base container under harmonic excitation along one of the base walls in vicinity of the sloshing lowest natural frequency. His approach was based on the perturbation method introduced by Narimanov [16] that assumed two steady-state sloshing regimes of planar waves and swirling wave. Due to embedded blunder found years later, this method failed to predict more sloshing regimes. As described by Bryant and Stiassnie [17], the following nonlinear regimes might take place in square-base container: planar two-dimensional wave, diagonal wave and swirling wave. They showed that near the primary resonance with respect to the fundamental sloshing mode, the two-dimensional regimes lose their stability for out-of-plane (or three dimensional) perturbation, whereas the three-dimensional regimes are stable. The same steady-state nonlinear sloshing regimes under longitudinal harmonic excitation were identified by Waterhouse and Ockendon [18], [19] in the vicinity of multi-modal nonlinear resonance. Based on Narimanov [16], Faltinsen et al. [11], [20]–[22] developed multimodal approach to investigate the nonlinear sloshing in a square-base container, using Bateman-Luke variational principle. This method yields multi-dimensional system of coupled ordinary differential equations of the time dependent multiple natural modes. A weakly nonlinear analysis near resonance of two dimensional ideal fluid horizontally forced rectangular tank waspresented by Forbes [23]. He introduced a novel spectral method which is a generalization of the technique presented by Faltinsen et al.[2]. This analytical method describes both regular and irregular liquid sloshing motion, and managed to obtain irregular solutions that were described by Chester and Bones [24], [25] and numerically by Frandsen [3]. In the work done by Ikeda et al. [26], [27] on nonlinear sloshing responses in square tank under obliquely horizontal excitation, super-position of two modes sloshing motion is combined to yield a swirling motion.

Due to the complex nature of the nonlinear sloshing dynamics and additional difficulty originating from the different tank shapes (i.e. container walls are not vertical and straight), a variety of numerical approached were developed. Three methods are widely used for sloshing simulations: finite-difference method (FDM, [28]), finite element method (FEM,[29]) and boundary element method (BEM, [30]). The latter is more recommended from reasons of computational complexity and time consumption [4]. Chen and Nokes [31] and Bredmose [32] observed irregular responses under horizontally excited rectangular tank using FDM. Liu and Lin [33] applied the volume of fluid (VOF) method to track free-surface movements under six degree of freedom containing rectangular tank motion. Their results for horizontal excitation were compared to linear analytical solution gave by Faltinsen [34], and three dimensional violent sloshing simulations were presented. Lohner et al. [35] VOF technique.

Koh et al. [36] developed a variationally coupled BEM-FEM for free-surface response and compared the results with experimental results for seismic excitation. Zhang [37] showed a semi-Lagrangian procedure applied to simulate fully-nonlinear sloshing waves in non-wall-sided tank (i.e. tank without vertical walls). His results were compared to open source CFD software OpenFOAM.

Due to the significant computational efforts required for simulation of the complete system, a number of approximate phenomenological models were developed in order to get at least qualitative insight into this phenomenon. A well-known approximate method for sloshing dynamical features imitation and resulting hydraulic forces estimation is equivalent mechanical model. Equivalent mechanical models for linear small-amplitude sloshing were based on pendulums and mass-spring systems were studied for different excitations and tank shapes [1], [38], [39]. However, since nonlinear sloshing regimes cannot be described by the linear models, major attempts were made to formulate nonlinear reduced order equivalent mechanical models [7], [8], [10], [26], [40], [41] that may predict and describe nonlinear responses observed experimentally. For instance, nonlinear rotary sloshing in a scale model of Centaur propellant tank at low fill level was modeled by combination of ordinary linear pendulum and a spherical pendulum (i.e. compound pendulum) by Kana [42]. In previous works, we treated hydraulic impacts regime with the help of reduced-order mechanical models. Both pendulum [10] and mass-spring systems [43], [44] were used to describe the vibro-impact regimes induced under a horizontal ground excitation. Tank elasticity and strongly nonlinear sloshing were considered, and different dynamical regimes were described with asymptotic tools that were validated numerically. In [44], analysis of realistic partially-filled tank was shown and mitigation of vibrations was optimized with the help tuned mass damper (TMD) and nonlinear energy sink (NES).

So far, detailed analytical explorations of liquid oscillations in a partially-filled vessel under horizontal harmonic ground excitation are limited to small-amplitude sloshing in rectangular and cylindrical vessels. We use the term “moderate” to denote the response regimes, in which the nonlinearity qualitatively modifies the responses, and leads to creation of the new ones, absent in the linear setting. However, the most violent nonlinear sloshing phenomena, such as impacts of the liquid applied to the vessel walls, are excluded. While being interesting and potentially dangerous, the moderate-amplitude liquid sloshing still lacks comprehensive analytic description. To get (albeit limited) insight into such regimes of the moderately nonlinear sloshing, we adopt and consider the equivalent mechanical model.. The moderate-nonlinearity is modeled by addition of a cubic spring to the linear stiffness; thus, the model is relevant both for linear moderately nonlinear sloshing regimes. Since the mass of the sloshing liquid in realistic conditions does not exceed 10% of the total mass of the system, the reduced-order model turns to be formally equivalent to well-studied oscillatory systems with cubic nonlinear energy sinks (NES) [45]–[48]. Conditions for existence and coexistence of periodic steady-state, weak-quasiperiodic and strongly modulated response (SMR) are obtained. The slow invariant manifold (SIM) describing the system slow flow-dynamics of 1:1 internal resonance is derived by multiple-scale analysis of the system. Analytical predictions are validated using numerical simulations.  Overview of different analytical, numerical and experimental methods is presented, and the sloshing regimes described by those methods are discussed. Finally, the results are compared qualitatively with previous experimental and computational data and show good agreement in terms of dynamical regimes.

# 2. Description of the model

## 2.1. Introducing the model.

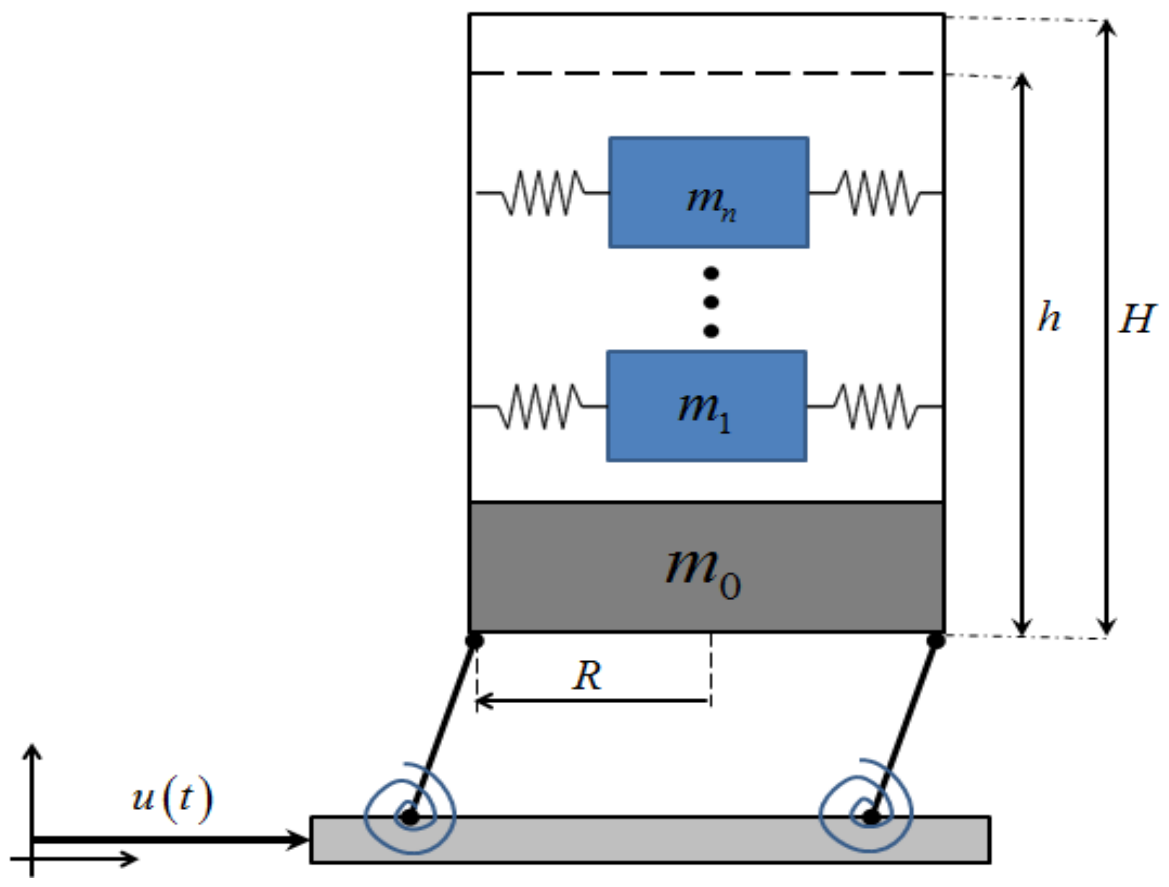

Figure 1-Scheme of multiple sloshing modes in partially-filled tank represented by a series of mass-spring systems interacting with structure walls.

Mechanical models of the liquid sloshing often use an infinite series of pendula or mass-spring systems to represent the free liquid surface oscillation using infinite series of sloshing modes. These models, primarily developed using linear sloshing theory, are shown for various types of tank geometries and excitation types, by Graham [49], Graham and Rodriguez [38] and others [1], [5], [39], [50]. Equivalent moment of inertia of a liquid in cylindrical containers has been estimated numerically by Partom ([51] and [52]) and verified experimentally by Werner and Coldwell) [53]. Parameters of equivalent pendulum, which corresponds to  the first asymmetric sloshing mode of cylindrical and rectangular tanks were studied by Dodge [39] and Abramson. As shown by Dodge [39], the mass of each modal pendulum decreases rapidly with increasing mode number (Figure 2). Consequently, it is reasonable to take into account only the first mode in the mechanical equivalent model, as long as the excitation frequency is far from the natural frequencies of the higher modes.

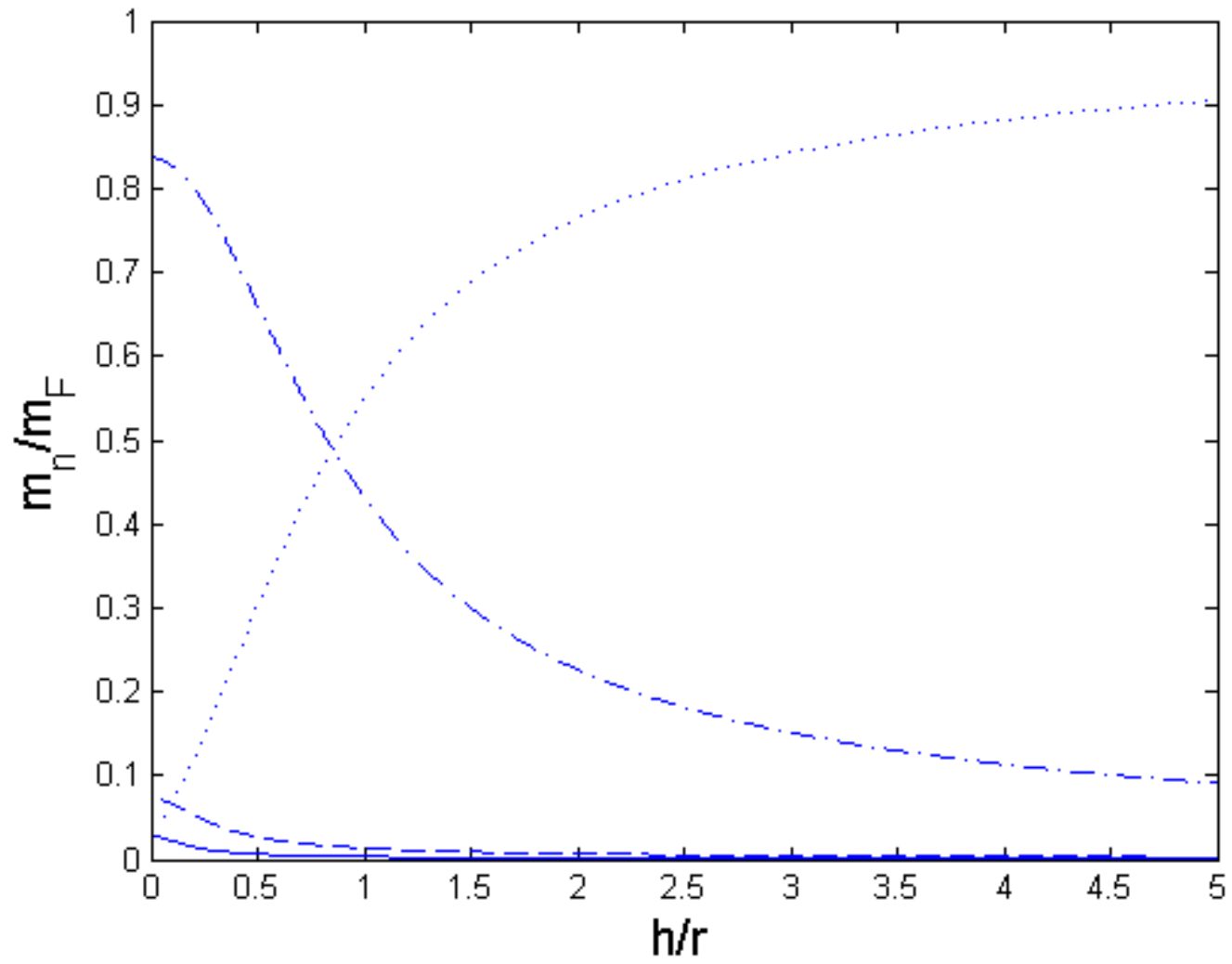

Figure 2-Ratios of the first three asymmetric sloshing modal masses $m_1$, $m_2$ and $m_3$ and fixed mass $m_0$ to the total fluid mass $m_F$ for cylindrical vessel; dotted-lime: $m_0/m_F$, dashed-dotted-line: $m_1/m_F$ ,dashed-line: $m_2/m_F$ solid-line: $m_3/m_F$ .

Extensive use was made for engineering purposes, for example to estimate the hydraulic loads exerted on tanker truck inner walls during breaking and turning and comparison the results with finite element simulation [54], and for resistance prediction of liquid filled vessels under earthquakes [55].

However, the ROMs based on linear sloshing theory suffer from a major shortcoming. Due to their linearity, they fail to describe nonlinear dynamical regimes, such as hydraulic impacts, rotary sloshing, amplitude jump phenomenon, nonlinear beatings, strongly modulated response, and nonlinear interaction between sloshing and structural modes.

The ROM presented in the current study purports to treat both linear sloshing oscillations, modeled by concentrated mass attached by linear spring to larger containing mass, and cubic nonlinearity representing both linear low-amplitude, and non-linear high-amplitude sloshing motion. It is noteworthy, that three dimensional sloshing regimes of rotary sloshing (i.e. swirling) is not addressed in the discussed model. The reason for this, beyond simplicity considerations, is that we primarily purported to describe analytically the resonant regime, which is approximately two- dimensional, and can be described as such.

Previous analytical studies indicate that response regimes in partially-filled tanks subjected to harmonic horizontal excitation are described by Duffing equation, which corresponds to weakly nonlinear behavior of mass-spring system [12]. Moreover, Moiseev showed that even though the asymptotic analysis treated steady-state solutions, comparison with experimental results show that its predictions are valid also for nonlinear transient waves description. An important conclusion of this analysis was that two dimensional liquid flow steady-state response changes from hard-spring to soft-spring response with increasing fluid water level. In rectangular tank with height $h$ and breadth $l$ , this transition occurs for critical normalized water level of $h^{*}/l \approx 0.3368$. Experiments made by Fultz [56] estimated this value to be $h^{*}/l = 0.28$ . According to Hermann and Timokha [57], this normalized critical water depth decreases monotonously with decreasing ratio of external forcing excitation amplitude and tank breadth, and the value obtained by Moiseev corresponds to the case for which this ratio tends to zero. The results of Hermann and Timokha were consistent with the experiments of Fultz for similar forcing amplitudes. Gu and Sethna [58], Gu et al. [59],Virnig et al. [60] have studied the effect of the liquid critical depth in rectangular tanks subjected to vertical sinusoidal excitation.

Following the arguments mentioned above, the model comprises linear oscillator subjected to harmonic external forcing, with internal particle with masses of *M* and *m*, respectively; where *M* is the total mass of the container and the non-sloshing liquid portion, and *m* is the liquid first sloshing mode mass. The internal particle (IP) is attached to the primary structure (PM) by both linear and cubic stiffnesses, where the natural frequency of the smaller mass is much lower than the structure's natural frequency, and for realistic cases of rather filled tanks, its mass is essentially smaller with respect to the mass of the containing structure as can be learned from Figure 2. Scheme of the system is presented in Figure 3. Absolute non-dimensional displacements of the primary mass and particle are denoted as $u(t)$ and $v(t)$ respectively.

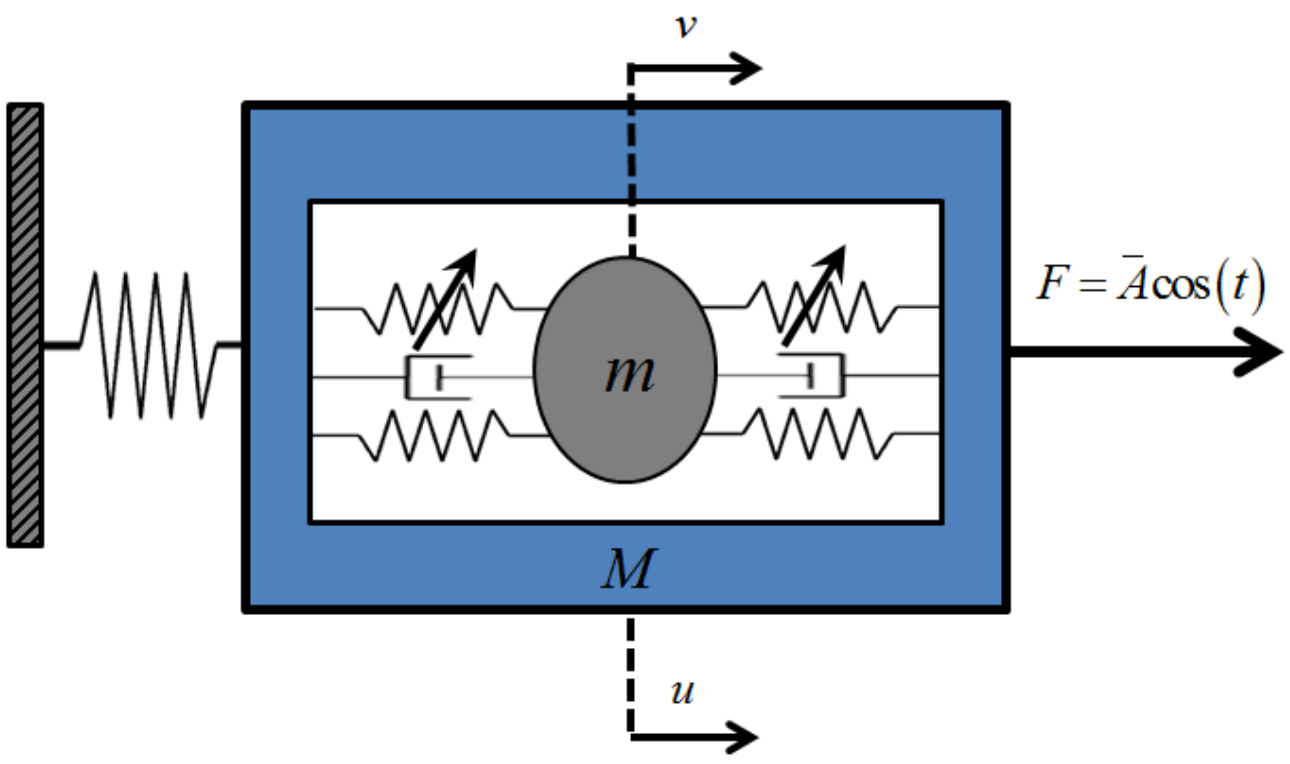

Figure 3-System scheme- linear oscillator as the primary system and internal particle with both linear and cubic attachments. The displacements of the primary mass and the impacting mass are denoted as u and v, respectively.

The IP is located inside a straight cavity in the PM, and in contrast to earlier explored vibri-impact systems [61] and vibro-impact NES [62], [63], it is attached to it through a linear viscosity $c$ and linear spring with stiffness $k_1$ and the mass of the first sloshing mode is *m*. The linear spring is required to mimic the small-amplitude oscillatory sloshing motion due to gravity. The additional cubic coupling is required to mimic moderate and high-amplitude sloshing, in accordance with earlier studies [8], [10], [40], which involve nonlinear characteristics associated with a cubic term. Moreover, cubic nonlinearity is the minimal-order symmetric nonlinearity, and therefore it is the obvious lowest-order correction of the linear coupling term. The external forcing is considered to be harmonic, with frequency $\Omega$ and amplitude $\bar{A}$. Hardening nonlinearity, corresponds to fluid level of $h < h*$ and positive value of parameter $k$ , whereas softening nonlinearity corresponds to $h > h*$ and negative value of parameter $k$ .

It is noteworthy that the framework of the presented asymptotic analysis is not limited to any specific tank shape. Any two dimensional tank shape with vertical walls in the free liquid surface zone can be treated. Moreover, we assume infinite roof height in order to eliminate both liquid spilling and interaction between the fluid and the tank roof.

## 2.2. Equations of motion

The equations of motion governing the liquid-filled vessel system are formulated with the help of Lagrange formalism. Kinetic and potential energies of the system are expressed as:

$$T = \frac{1}{2} M u_{tt}^2 + \frac{1}{2} m v_{tt}^2$$
$$V = \frac{1}{2} k_1 u^2 + \frac{1}{2} k_2 (v-u)^2 + \frac{1}{4} \bar{k} (v-u)^4 \tag{1}$$

Hence, the Lagrangian is given by the following expression:

$$L = T - V = \frac{1}{2} M u_{tt}^2 + \frac{1}{2} m v_{tt}^2 - \frac{1}{2} k_1 u^2 - \frac{1}{2} k_2 (v-u)^2 - \frac{1}{4} \bar{k} (v-u)^4 \tag{2}$$

Damping is taken into account through Rayleigh's dissipation function:

$$D = \frac{1}{2} c \left( v_t - u_t \right)^2 \tag{3}$$

Thus, by using the extended Hamilton's principle, the dimensional equations of motion are written as follows:

$$M u_{tt} + k_1 u + k_2 (u-v) + c(\dot{u} - \dot{v}) + \bar{k} (u-v)^3 = \bar{A} \cos(\Omega t)$$
$$m v_{tt} + k_2 (v-u) + c(\dot{v} - \dot{u}) + \bar{k} (v-u)^3 = 0 \tag{4}$$

We apply simple manipulations and time normalization $\tau = \Omega t$ on system (4) to yield the following normalized non-dimensional equations of motion:

$$\ddot{u} + (1 + \varepsilon\sigma) u + \varepsilon^2 \beta (u-v) + \varepsilon\lambda (\dot{u} - \dot{v}) + \varepsilon k (u-v)^3 = \varepsilon A \cos(\tau)$$
$$\ddot{v} + \varepsilon\beta (v-u) + \lambda (\dot{v} - \dot{u}) + k (v-u)^3 = 0 \tag{5}$$

Here dots represent differentiation with respect to the normalized time $\tau$, and the normalization and non-dimensional parameters governing the system dynamics are as follows:

$$\omega_1^2 = \frac{k_1}{M}, \omega_2^2 = \frac{k_2}{m}, \Omega = \frac{\omega_1}{\sqrt{1+\varepsilon\sigma}}, \varepsilon A = \frac{\bar{A}}{M\Omega^2}$$
$$\varepsilon\beta = \left( \frac{\omega_2}{\omega_1} \right)^2 (1 + \varepsilon\sigma), k = \frac{\bar{k}}{m\Omega^2}, \lambda = \frac{c}{m\Omega} \tag{6}$$

The following coordinate transformation is used:

$$X(t) = u(t) + \varepsilon v(t)$$
$$w(t) = u(t) - v(t) \tag{7}$$

Here $w$ the IP relative displacement with respect to the PM, and coordinate $X$ is proportional to the displacement of the center-of-mass. Following [63], [64], the transformed non-dimensional equations of motion:

$$\begin{aligned} &\ddot{X}+\frac{1+\varepsilon\sigma}{1+\varepsilon}X+\frac{\varepsilon\left(1+\varepsilon\sigma\right)}{1+\varepsilon}w=\varepsilon A\cos\left(t\right)\\ &\ddot{w}+\frac{1+\varepsilon\sigma}{1+\varepsilon}X+\varepsilon\eta w+\left(1+\varepsilon\right)\lambda\dot{w}+k\left(1+\varepsilon\right)w^{3}=\varepsilon A\cos\left(t\right) \end{aligned} \tag{8}$$

Here $\eta=\frac{1+\varepsilon\sigma}{1+\varepsilon}+\beta\left(1+\varepsilon\right)$. In the following section, weakly nonlinear slashing periodic regimes will be described analytically.

# 3. Dynamical regimes

## 3.1. Periodic regimes

The following complex variables are introduced:

$$\begin{aligned} &\dot{X}+iX=\varphi_{1}e^{it}\\ &\dot{w}+iw=\varphi_{2}e^{it} \end{aligned} \tag{9}$$

We substitute (9) into system (8) to obtain the following slow-flow system:

$$\begin{aligned} &\dot{\varphi}_{1}+\frac{i\varepsilon\left(1-\sigma\right)}{2\left(1+\varepsilon\right)}\left(\varphi_{1}+\bar{\varphi}_{1}e^{-2it}\right)-\frac{i\varepsilon\left(1+\varepsilon\sigma\right)}{2\left(1+\varepsilon\right)}\left(\varphi_{2}-\bar{\varphi}_{2}e^{-2it}\right)=\frac{\varepsilon A}{2}\left(1-e^{-2it}\right)\\ &\dot{\varphi}_{2}+\frac{i}{2}\left(1-\varepsilon\eta\right)\left(\varphi_{2}-\bar{\varphi}_{2}e^{-2it}\right)+\frac{\lambda\left(1+\varepsilon\right)}{2}\left(\varphi_{2}+\bar{\varphi}_{2}e^{-2it}\right)-\frac{i\left(1+\varepsilon\sigma\right)}{2\left(1+\varepsilon\right)}\left(\varphi_{1}-\bar{\varphi}_{1}e^{-2it}\right)\\ &\quad+\frac{ik\left(1+\varepsilon\right)}{8}\left(\varphi_{2}e^{it}+\bar{\varphi}_{2}e^{-it}\right)^{3}e^{-it}=\frac{\varepsilon A}{2}\left(1-e^{-2it}\right) \end{aligned} \tag{10}$$

Here the bar denotes complex conjugate. The problem analyzed is response regime of system (10) in vicinity of both primary and internal resonances of 1:1. Thus, we may assume that the complex modulation variables $\varphi_1$ and $\varphi_2$ are slow with respect to the external forcing. Thus, by averaging with respect to the fast time scale the approximate averaged system is as follows:

$$\begin{aligned}
&\varphi_1' + \frac{i\varepsilon(1-\sigma)}{2(1+\varepsilon)}\varphi_1 - \frac{i\varepsilon(1+\varepsilon\sigma)}{2(1+\varepsilon)}\varphi_2 = \frac{\varepsilon A}{2} \\
&\varphi_2' + \frac{i}{2}(1-\varepsilon\eta)\varphi_2 + \frac{\lambda(1+\varepsilon)}{2}\varphi_2 - \frac{i(1+\varepsilon\sigma)}{2(1+\varepsilon)}\varphi_1 - \frac{3ki(1+\varepsilon)}{8}|\varphi_2|^2\varphi_2 = \frac{\varepsilon A}{2}
\end{aligned} \tag{11}$$

Differentiation with respect to time is denoted by apostrophe in order distinguish it from the original fast time scale of the system. The averaging method cannot be considered as rigorous due to its deviation from the averaging theorem validity. However, it was successfully used in many previous studies [41], [65].

Periodic solutions of system (8) correspond to fixed points of system (11). Thus, we eliminate the derivatives to yield the following algebraic equations:

$$\begin{aligned}
&\frac{i\varepsilon(1-\sigma)}{2(1+\varepsilon)}\varphi_{10} - \frac{i\varepsilon(1+\varepsilon\sigma)}{2(1+\varepsilon)}\varphi_{20} = \frac{\varepsilon A}{2} \\
&\frac{i}{2}(1-\varepsilon\eta)\varphi_{20} + \frac{\lambda(1+\varepsilon)}{2}\varphi_{20} - \frac{i(1+\varepsilon\sigma)}{2(1+\varepsilon)}\varphi_{10} - \frac{3ki(1+\varepsilon)}{8}|\varphi_{20}|^2\varphi_{20} = \frac{\varepsilon A}{2}
\end{aligned} \tag{12}$$

Here $\varphi_{10}$ and $\varphi_{20}$ are the steady-state values of the complex modulation functions, i.e. at the fixed points. System (12) can be solved easily to yield the following expression for steady-state solution amplitude:

$$\begin{aligned}
&N_0^2\left(a_1^2 + \left(a_2 + a_3 N_0^2\right)^2\right) = a_4^2 A^2 \\
&\varphi_{20} = N_0 e^{i\theta_0}
\end{aligned} \tag{13}$$

By taking $Z = N_0^2$, we bring system (13) to an equivalent form:

$$\alpha_3 Z_0^3 + \alpha_2 Z_0^2 + \alpha_1 Z_0 + \alpha_4 = 0 \tag{14}$$

Expressions for coefficients $\alpha_i$ are given in Appendix A.

Existence of multiple periodic solutions may take place for different parameter values. Transition from qualitatively different periodic solutions corresponds to local bifurcation of the periodic solutions.

### 3.1.1. Saddle-Node bifurcation

In this part, we will find the boundary which separated between single-periodic solution zone and three periodic solutions zone in the parameter space. These transitions correspond to saddle-node (SN) bifurcation in equation (14). The boundary is obtained by the necessary and complete condition for SN bifurcation of equating the derivative of equation (14) to zero:

$$3\alpha_3 Z_0^2 + 2\alpha_2 Z_0 + \alpha_1 = 0 \tag{15}$$

Using polynomials (14) and (15), we obtain the following expression for the SN boundary:

$$3\alpha_3\left(\alpha_1\alpha_2 - 9\alpha_3\alpha_4\right)^2 + 2\alpha_2\left(\alpha_1\alpha_2 - 9\alpha_3\alpha_4\right)\left(6\alpha_1\alpha_3 - 2\alpha_2^2\right) + \alpha_1\left(6\alpha_1\alpha_3 - 2\alpha_2^2\right)^2 = 0 \tag{16}$$

Equation (16) corresponds to nullifying of the discriminants of cubic polynomial (14) .For a certain parameter set of $\varepsilon, \sigma, k$ the SN bifurcation, described by equation (16), is represented by a contour on plane $(\mathrm{A}, \lambda)$, as shown in Figure 4. The single solution zone and three solutions zones correspond to the cases of negative and positive values of the discriminant, respectively.

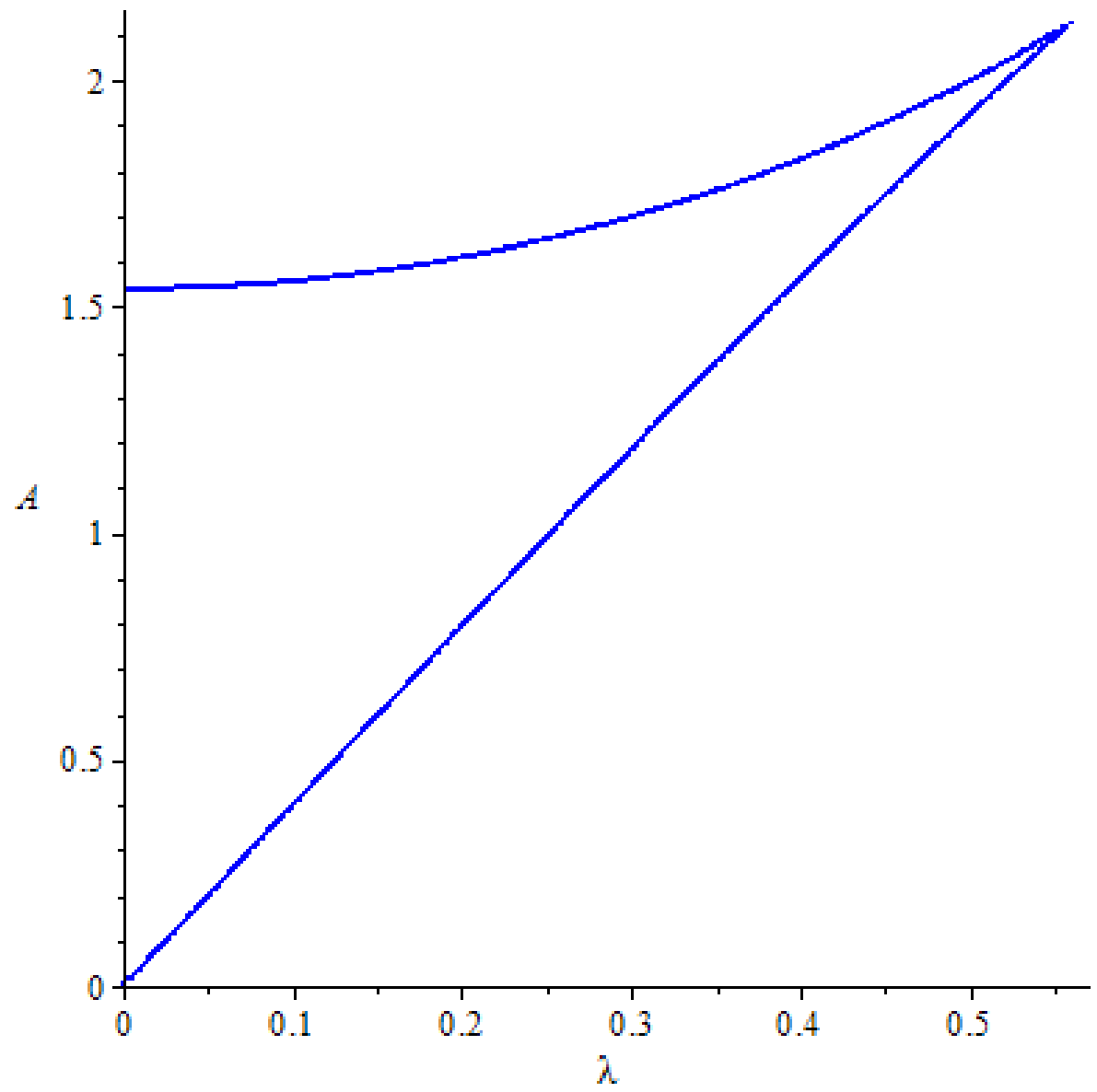

Figure 4-Projection of the saddle-node bifurcation for $\varepsilon = 0.05, k = 4/3, \beta = 5, \sigma = 5$

We substitute expressions $\alpha_i$ into equation (16) and obtain polynomial in $A$ or $\lambda$. If we take A as a bifurcation variable and $\lambda$ as a parameter, we can calculate the critical maximum value of $\lambda$ for SN bifurcation by eliminating the discriminant of equation (16) as a polynomial in A.

$$\lambda_{SN,cr} = \frac{1}{\sqrt{3}}\left|\frac{\sigma}{1-\sigma} + \varepsilon\beta\right| \tag{17}$$

The maximum value of A is calculated by eliminating the discriminant of equation (16) as a polynomial in $\lambda$, and taking the higher value obtained:

$$A_{SN,cr} = \frac{4}{9}\sqrt{\frac{2}{\left|k\left(\sigma-1\right)\right|}}\left(\sigma-\varepsilon\beta\left(\sigma-1\right)\right)^{3/2} \quad (18)$$

As one can see from equation (18), necessary condition for existence of SN bifurcation is positive value under the square root. This condition yields the following quadratic polynomial inequality in $\sigma$:

$$k\left(\bar{\beta}-1\right)\sigma^2+k\left(1-2\bar{\beta}\right)+k\bar{\beta}<0 \quad (19)$$

Here $\bar{\beta}=\varepsilon\beta$ . The roots of polynomial (19) are as follows:

$$\sigma_{1,2}=1+\frac{1\pm\operatorname{sgn}\left(k\right)}{2\left(\bar{\beta}-1\right)}\Rightarrow\sigma_1=1,\ \sigma_2=\frac{\bar{\beta}}{\bar{\beta}-1} \quad (20)$$

For positive values of $k$, SN bifurcations are obtained for $1<\sigma<\bar{\beta}/\left(\bar{\beta}-1\right)$ if $\bar{\beta}>1$, and for $\sigma<\bar{\beta}/\left(\bar{\beta}-1\right)\, or\, \sigma>1$ if $0<\bar{\beta}<1$. For negative values of $k$, SN bifurcations are obtained for $\bar{\beta}/\left(\bar{\beta}-1\right)<\sigma<1$ if $0<\bar{\beta}<1$, and for $\sigma<0\ or\,\sigma>\bar{\beta}/\left(\bar{\beta}-1\right)$ if $\bar{\beta}>1$.

For the special case of $\bar{\beta}=0$ and $k>0$, described by Starosvetsky and Gendelman [48], SN bifurcations are obtained for $\sigma<0$ or $\sigma>1$. Isocline maps of $A_{cr}$ and $\lambda_{cr}$ versus $\sigma$ and $\bar{\beta}$ for $k = 4/3$ are shown in

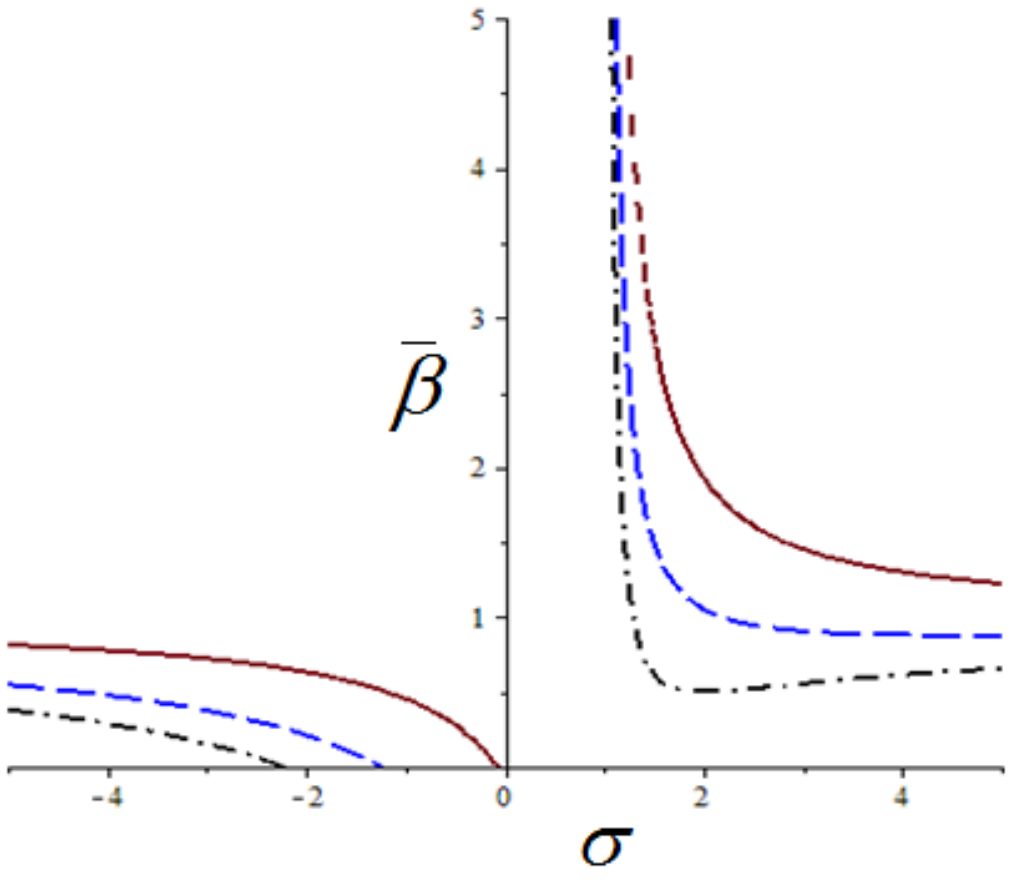

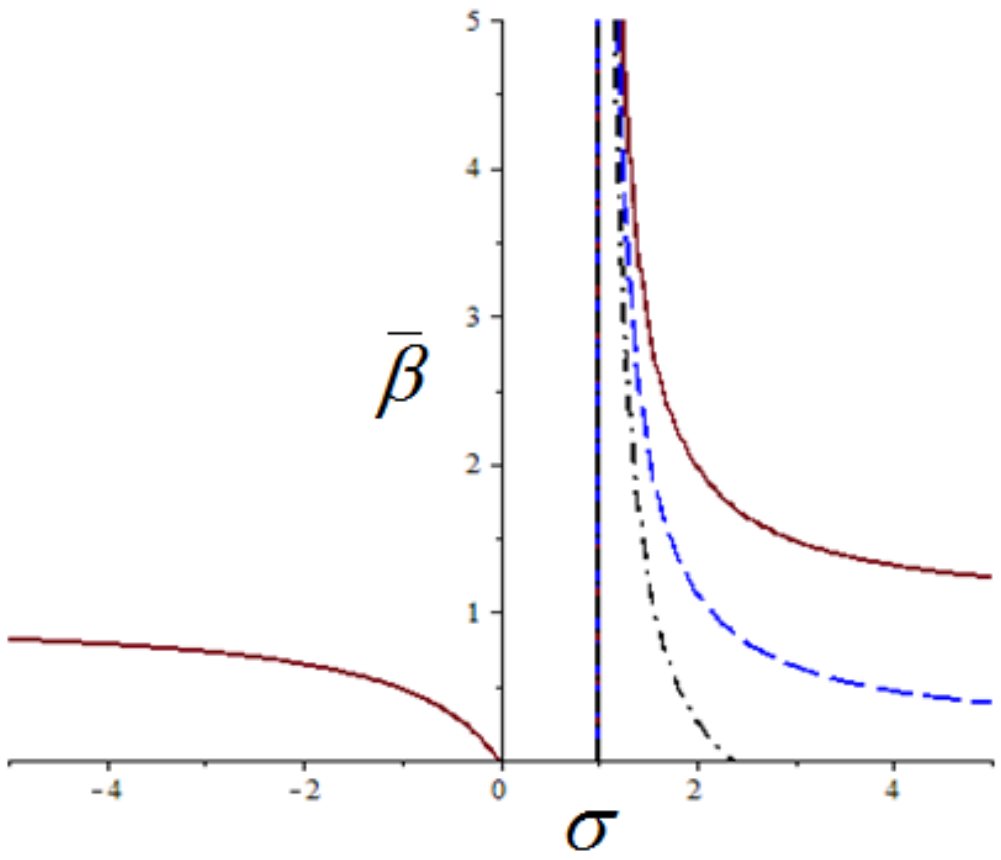

Figure 5(a) and (b), respectively.

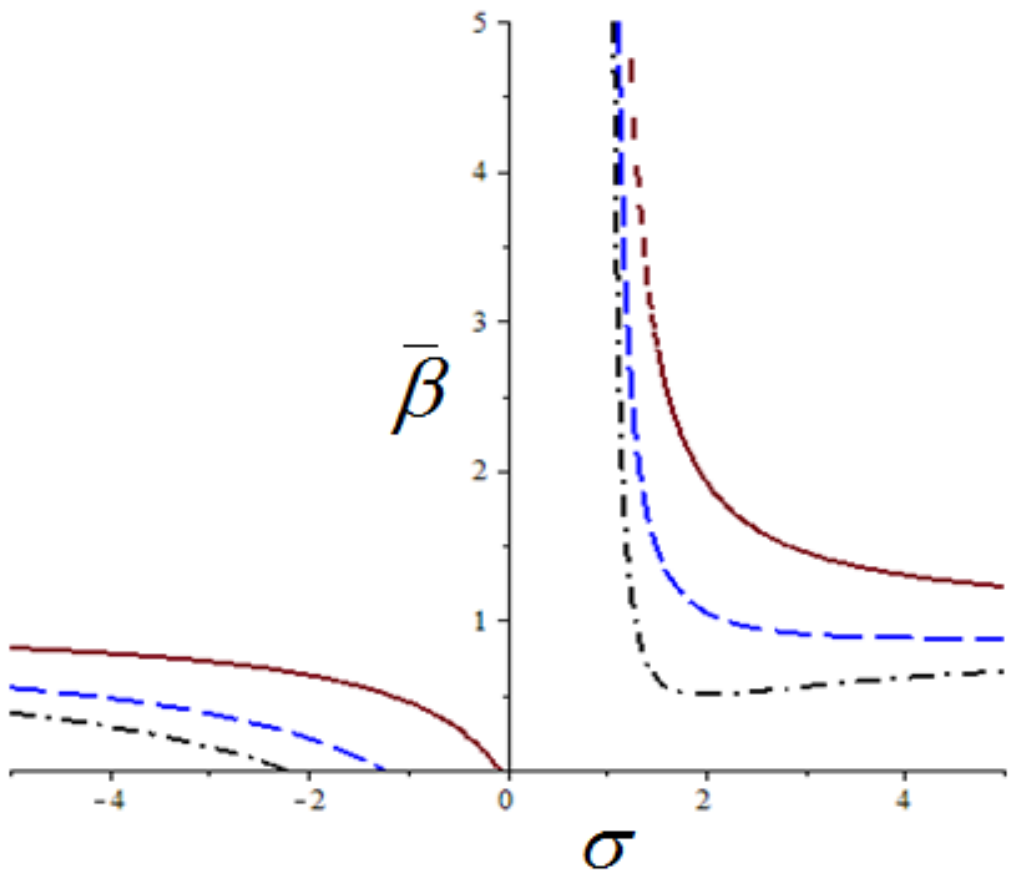

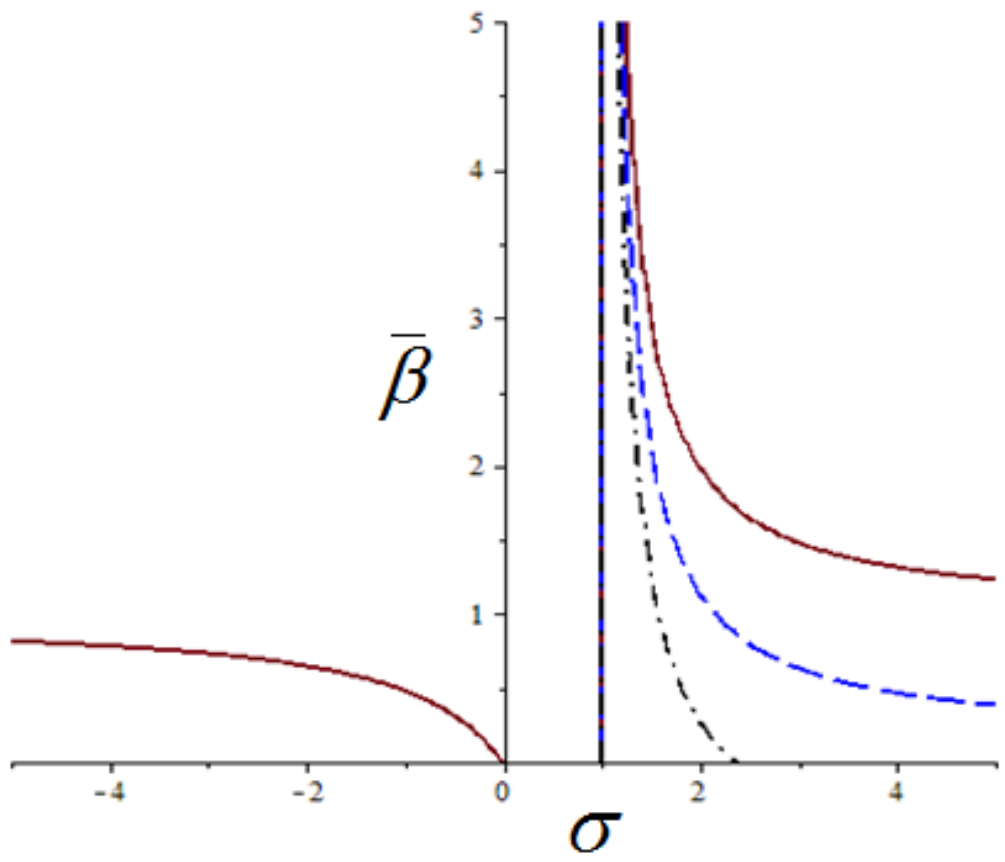

Figure 5- Isocline maps of $A_{cr}$ (left) and $\lambda_{cr}$ (right) for existence of SN bifurcation versus $\sigma$ and $\bar{\beta}$ for $k = 4/3$; left: solid-red: $A_{SN,cr} = 0.01$, dashed-blue: $A_{SN,cr} = 0.5$, dashed-dot: $A_{SN,cr} = 1$. Right: solid-red: $\lambda_{SN,cr} = 0.01$, dashed-blue: $\lambda_{SN,cr} = 0.2$, dashed-dot: $\lambda_{SN,cr} = 0.5$

## 3.2. Quasiperiodic regimes

In the following section, we refer to two types of quasiperiodic responses; the first one is weak quasiperiodic response, which corresponds to loss of stability of the periodic regime and formation of the limit cycle through Hopf bifurcation. The second quasiperiodic regime is associated with relaxation oscillations, which are characterized by alternating fast and slow response amplitude variations. Detailed investigation of those regimes is described by Starosvetsky and Gendelman [46], [66].

### 3.2.1. Weak quasiperiodic responses

In order to obtain the conditions for Hopf bifurcation, we perform small perturbation around the fixed points of system (11):

$$\begin{aligned} \varphi_1 &= \varphi_{10} + \delta_1 \\ \varphi_2 &= \varphi_{20} + \delta_2 \end{aligned} \tag{21}$$

Here $\delta_1$ and $\delta_2$ are small perturbation variables, of order $\varepsilon$. We substitute equation (21) into the averaged system (11) and keep merely the linear terms to yield the following linearized system, shown in its matrix form:

$$\begin{pmatrix} \delta_1' \\ \bar{\delta}_1' \\ \delta_2' \\ \bar{\delta}_2' \end{pmatrix} = \begin{pmatrix} -c_{11} & 0 & c_{12} & 0 \\ 0 & c_{11} & 0 & -c_{12} \\ c_{21} & 0 & -(c_{22}+c_{23})+2c_{24}\left|\varphi_{20}\right|^2 & c_{24}\varphi_{20}^2 \\ 0 & -c_{21} & -c_{24}\bar{\varphi}_{20}^2 & -(c_{22}-c_{23})-2c_{24}\left|\varphi_{20}\right|^2 \end{pmatrix} \begin{pmatrix} \delta_1 \\ \bar{\delta}_1 \\ \delta_2 \\ \bar{\delta}_2 \end{pmatrix} \tag{22}$$

Expressions for coefficients $c_{ij}$ are given in Appendix B. The corresponding characteristic polynomial of system (22) is as follows:

$$P(\mu) = \mu^4 + \gamma_1\mu^3 + \gamma_2\mu^2 + \gamma_3\mu + \gamma_4 \tag{23}$$

Expressions for coefficients $\gamma_i$ are given in Appendix C. Hopf bifurcation corresponds to $\mu = \pm i\Omega_H$, $\Omega_H \in \mathbb{R}$. That implies $\Omega_H^2 = \gamma_1/\gamma_3$ and $\gamma_3^2/\gamma_1^2 - \gamma_2\gamma_3/\gamma_1 + \gamma_4 = 0 \Rightarrow \Delta_3 = 0$. Thus, the condition for Hopf bifurcation and Hopf frequencies are as follows:

$$\begin{aligned} &\gamma_3^2 - \gamma_1\gamma_2\gamma_3 + \gamma_4\gamma_1^2 = 0 \\ &\Omega_H = \pm\frac{1}{2}\sqrt{\frac{\varepsilon(1+\varepsilon\sigma)}{1+\varepsilon}} \end{aligned} \tag{24}$$

Locus of SN and Hopf bifurcations on $(\lambda, A)$ plane for several values of $\beta$ is shown in Figure 6.

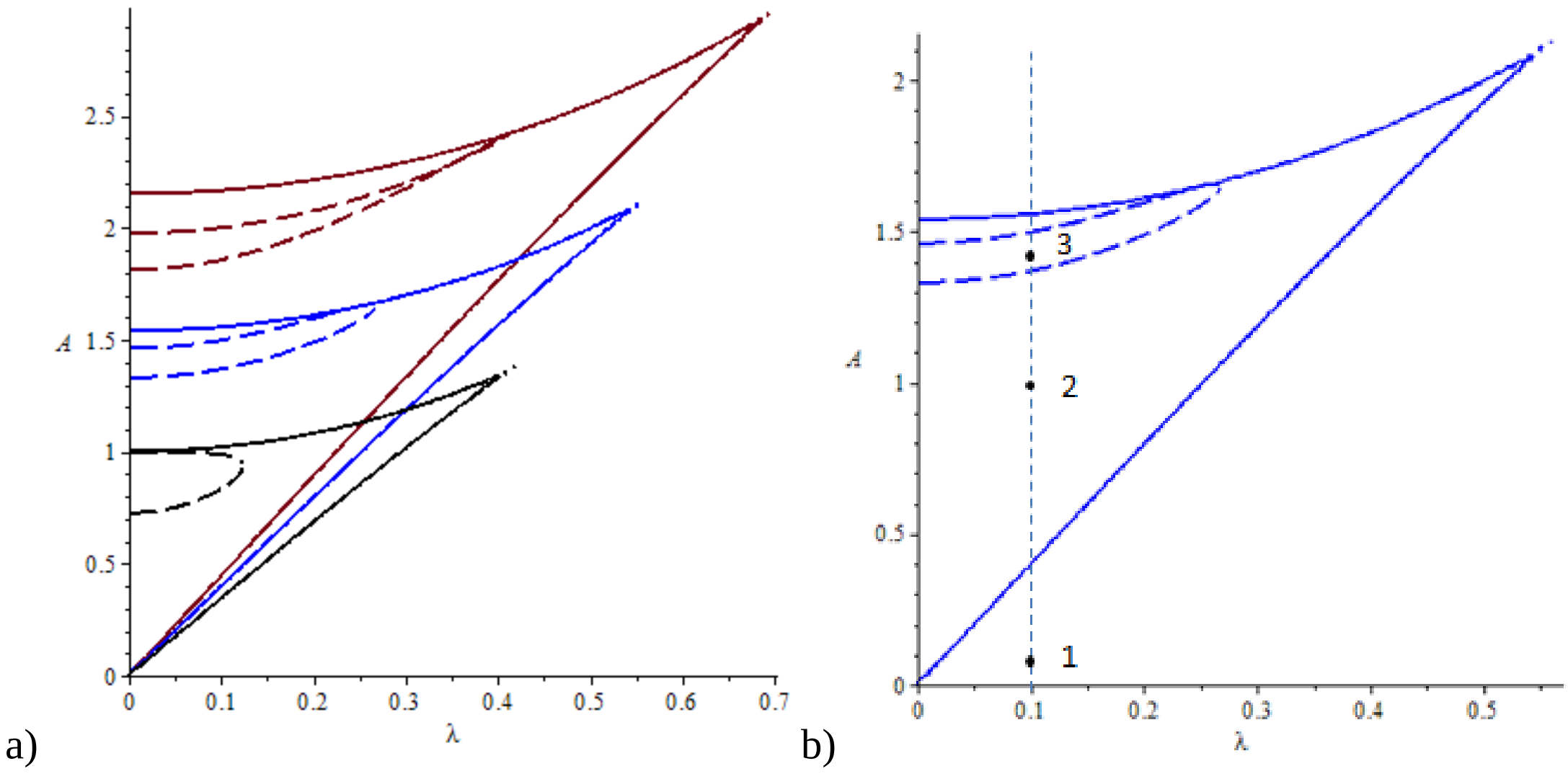

Figure 6- Projection of the saddle-node and Hopf bifurcations for: $k = 4/3, \sigma = 5, \varepsilon = 0.05$; SN bifurcations: solid line, Hopf bifurcation: dashed line. a) Red: $\beta = 0$, blue: $\beta = 5$, black: $\beta = 10$ (color online); b) for $\beta = 5$, point 1: $A = 0.1$ existence of periodic solutions; point2: $A = 1$ coexistence of both weakly and strongly quasiperiodic regimes; point 3: $A = 1.4$ coexistence of both periodic and strongly quasiperiodic regimes.

As one can see in Figure 6 (b), the bifurcations locus consists of three different zones: point 1: existence of periodic solutions; point2: coexistence of both weakly and strongly quasiperiodic regimes; and point 3: coexistence of both periodic and strongly quasiperiodic regimes.

Equation (24) is a bi-quadratic polynomial with respect to *N*, and has two roots. Critical damping value for existence of Hopf bifurcation $\lambda_{H,cr}$ is obtained by equating both of them:

$$\lambda_{H,cr} = \frac{1}{\sqrt{3}} \left| \bar{\beta} + \frac{\varepsilon\sigma - 1}{(1+\varepsilon)} \right| \tag{25}$$

Isocline maps of $\lambda_{cr}$ versus $\sigma$ and $\bar{\beta}$ for $\varepsilon = 0.05$ are shown in Figure 7.

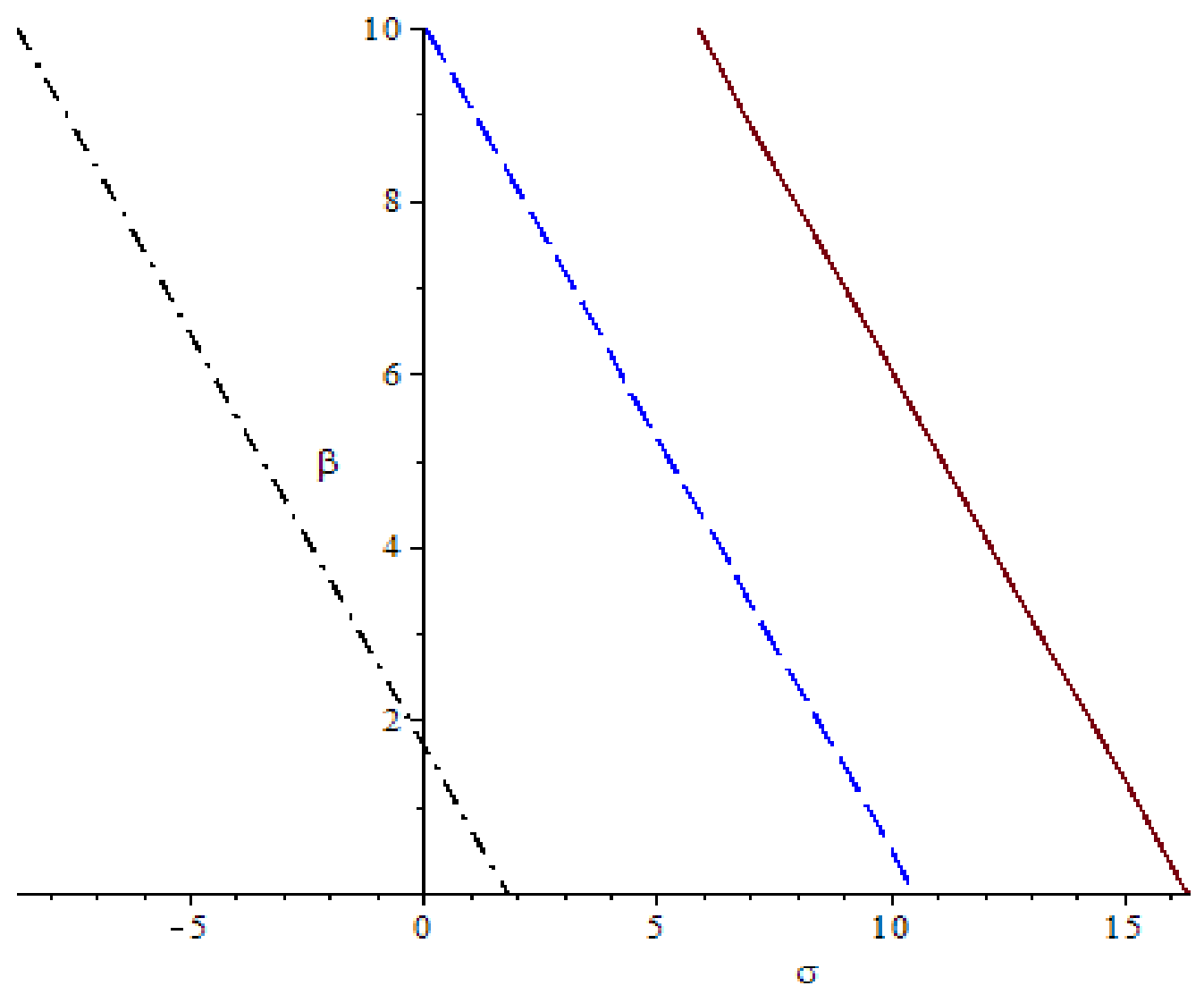

Figure 7- Isocline maps of $\lambda_{cr}$ versus $\sigma$ and $\bar{\beta}$ for $\varepsilon = 0.05$. Solid-red: $\lambda_{H,cr} = 0.1$, dashed-blue: $\lambda_{H,cr} = 0.268$, dash-dotted black: $\lambda_{H,cr} = 0.5$.

As one can see in Figure 6, each projection of bifurcation manifold of $\beta = 5$ consists of three different zones, corresponding to points 1, 2 and 3. Hopf and SN bifurcations diagrams for fixed values of parameters corresponding to the three points are presented in a)

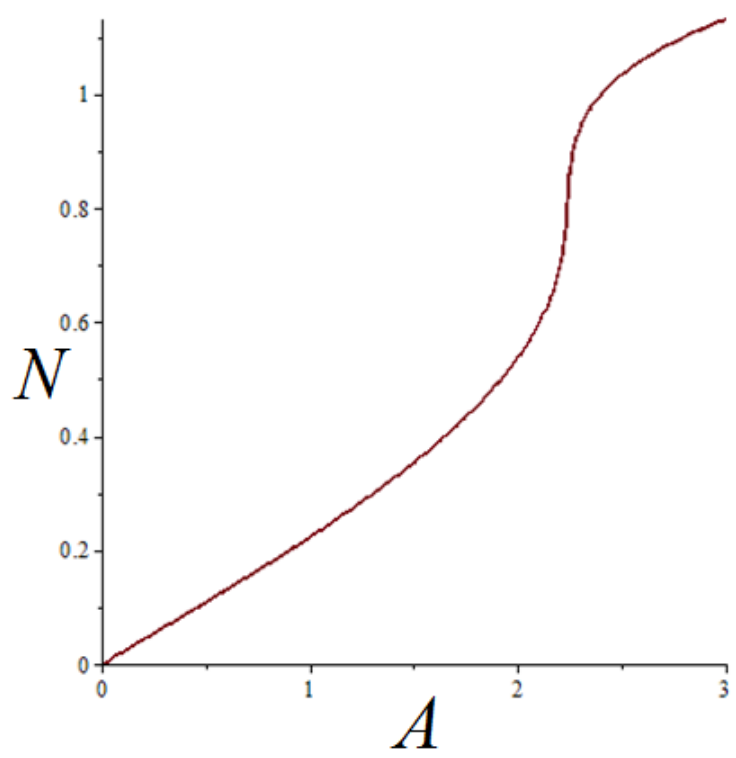

b)

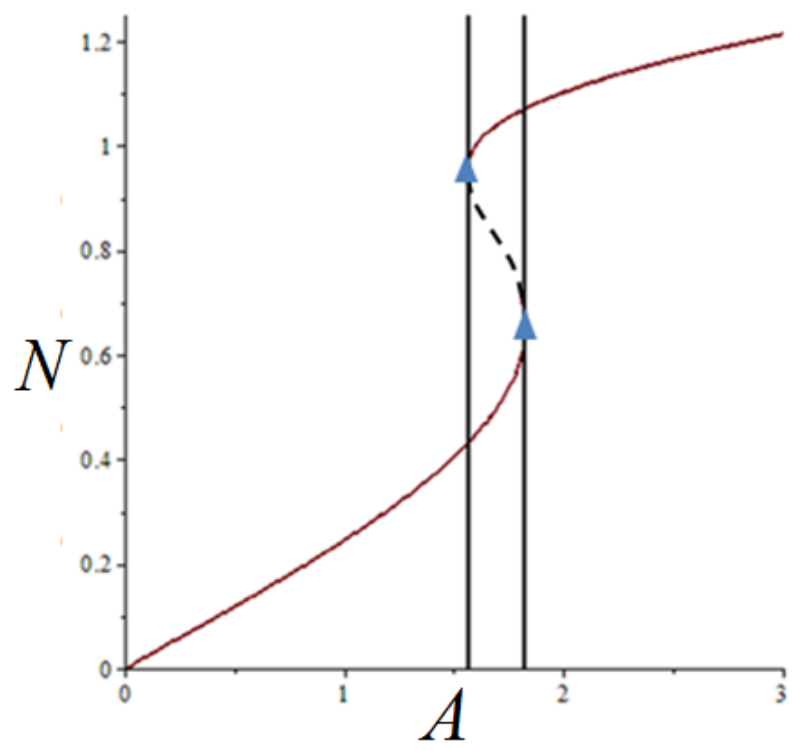

c)

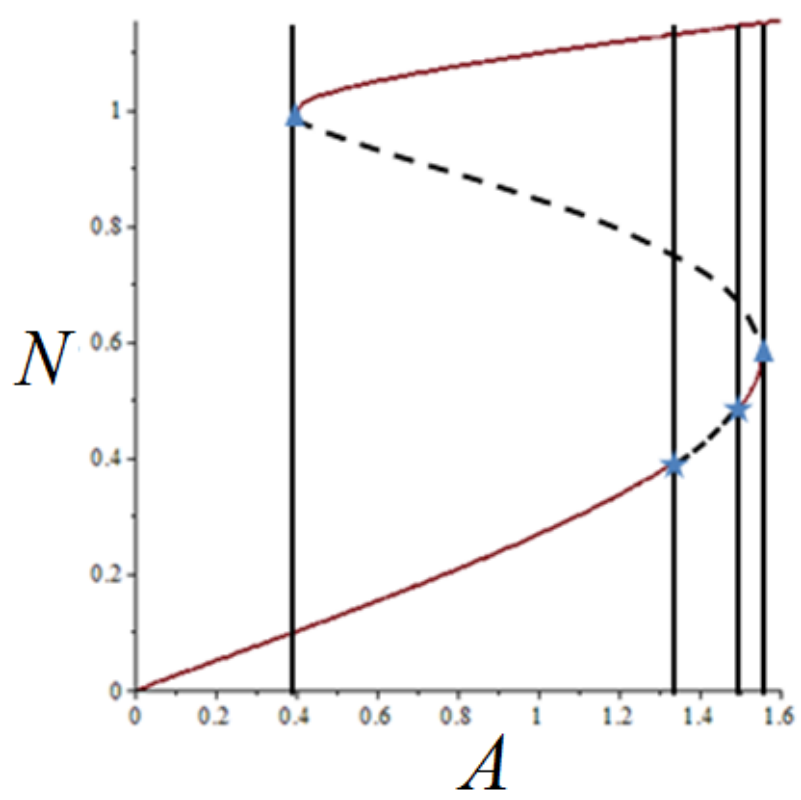

Figure 8 (a-c).

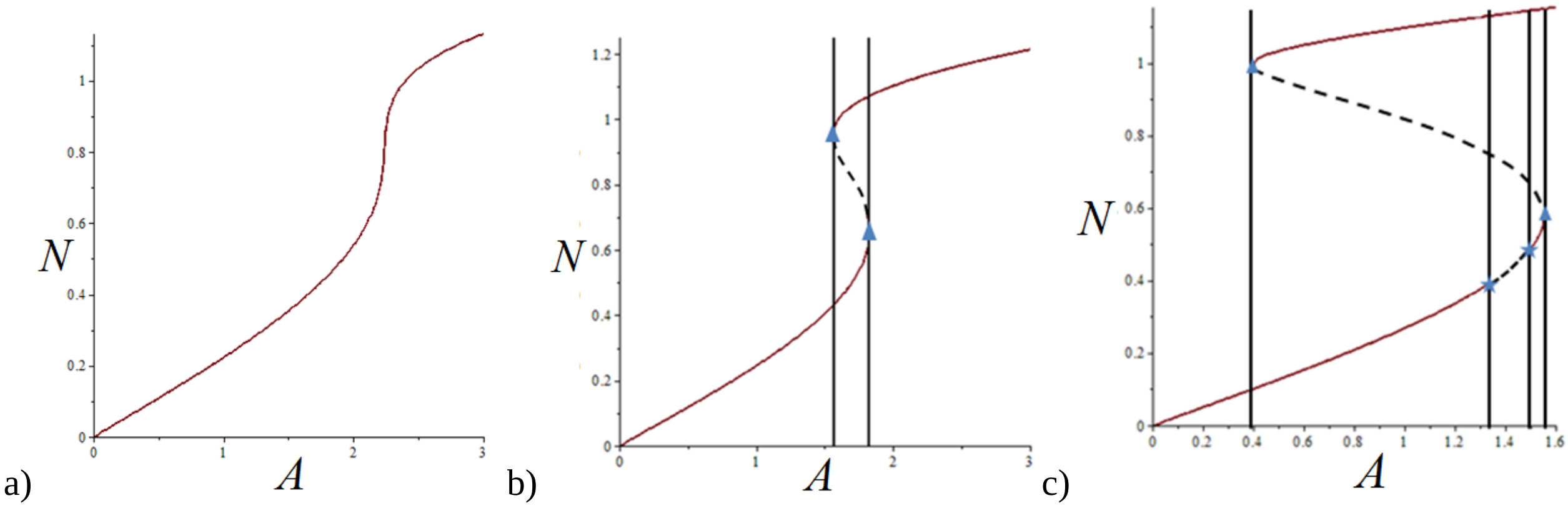

Figure 8- SN bifurcations (marked by traingles) and Hopf bifurcations (marked by stars) diagram for: $k = 4/3, \sigma = 5, \varepsilon = 0.05$, $\beta = 5 \Rightarrow \lambda_{cr,H} = 0.268$, $\lambda_{cr,SN} = 0.577$. Stable solutions- solid lines, unstable solutions- dashed line; a) $\lambda = 0.6 > \lambda_{cr,SN}$; b) $\lambda_{cr,H} < \lambda = 0.4 < \lambda_{cr,SN}$; c) $\lambda = 0.1 < \lambda_{cr,H}$.

### 3.2.2. Strongly Modulated Responses (SMR)

The regions of stability of periodic solutions discussed previously were analyzed by tools of local bifurcations and perturbation methods. Those methods cannot be used for description of the additional existing regime of SMR. As shown in previous studies [46], [48], [66], this regime is characterized by aggressive amplitude modulations and exists in vicinity of 1:1 resonance. In contrary to the current study, those previous studies focused on essentially nonlinear attachment, known as cubic nonlinear energy sink (NES), and its contribution in terms of mitigation performances of linear oscillator. In this section, following Starosvetsky and Gendelman [48], we obtain the conditions for existence of SMR using analytical approach of multiple-scales.

By differentiation and simple algebraic procedures, we reduce system (11) to the following single second-order equations with respect to $\varphi_2$:

$$\frac{d^2}{dt^2}\varphi_2+\frac{d}{dt}\left(\bar{c}_1\varphi_2+\bar{c}_2\left|\varphi_2\right|^2\varphi_2\right)-\left(\bar{c}_3\varphi_2+\bar{c}_4\left|\varphi_2\right|^2\varphi_2+\frac{\varepsilon A}{2}\bar{c}_5\right)=0 \tag{26}$$

Expressions for coefficients $\bar{c}_i$ are given in Appendix D. Multiple-scales approach is introduced:

$$\begin{aligned}&\varphi_2=\varphi_2\left(T_0,T_1\right)\\&T_n=\varepsilon^n t,\ n=0,1,\ldots\\&\frac{d}{dt}=D_0+\varepsilon D_1\end{aligned} \tag{27}$$

Here $\varepsilon$ is a small parameter and $D_i=\dfrac{\partial}{\partial T_i}$ , $D_i^2=\dfrac{\partial^2}{\partial T_i^2}$. Substituting equation (27) into equation (26)and collecting equal orders terms with respect to small parameter $\varepsilon$ obtains the following expressions:

$$\begin{aligned}&\mathbf{O}(\mathbf{1}):D_0^2\varphi_2+D_0\left[\left(\frac{\lambda}{2}+\frac{i}{2}\right)\varphi_2-\frac{3ki}{8}\left|\varphi_2\right|^2\varphi_2\right]=0\\&\mathbf{O}(\boldsymbol{\varepsilon}):2D_0D_1\varphi_2+D_0\left[\left(\frac{i(1-\sigma)}{2}+\frac{\lambda}{2}-\frac{i}{2}(1+\beta)\right)\varphi_2-\frac{3ki}{8}\left|\varphi_2\right|^2\varphi_2\right]\\&\quad+D_1\left[\left(\frac{\lambda}{2}+\frac{i}{2}\right)\varphi_2-\frac{3ki}{8}\left|\varphi_2\right|^2\varphi_2\right]-\left(-\frac{1}{4}\left(1+i(1-\sigma)(\lambda+i)\right)\varphi_2-\frac{3k(1-\sigma)}{16}\left|\varphi_2\right|^2\varphi_2+\frac{Ai}{4}\right)=0\end{aligned} \tag{28}$$

The solution of the first equation in system(28) represents the fast evolution of the averaged system. Integration of this equation with respect to time scale $T_0$ yields the following:

$$D_0\varphi_2+\left[\left(\frac{\lambda}{2}+\frac{i}{2}\right)\varphi_2-\frac{3ki}{8}\left|\varphi_2\right|^2\varphi_2\right]=C\left(T_1\right) \tag{29}$$

Here $C\left(T_1\right)$ is an arbitrary function of slow time scale $T_1$, i.e. constant with respect to time scale $T_0$. Stationary points $\Phi\left(T_1\right)$ of equation (29) correspond to elimination of the derivative, and obtained by the following algebraic equation:

$$\left[\left(\frac{\lambda}{2}+\frac{i}{2}\right)\Phi-\frac{3ki}{8}\left|\Phi\right|^2\Phi\right]=C\left(T_1\right) \tag{30}$$

We substitute the polar transformation $\Phi\left(T_1\right)=N\left(T_1\right)e^{i\theta\left(T_1\right)}$ into equation(30), and use several simple manipulations:

$$Z\left(\lambda^2+\left(1-\frac{3k}{4}Z\right)^2\right)=4\left|C\left(T_1\right)\right|^2 \tag{31}$$

Here $Z\left(T_1\right)=N\left(T_1\right)^2$. As one can see, equation (31) may have either a single solution or three solutions. The former case corresponds to monotony of the function in the left-hand side of equation (31), and the latter corresponds to existence of extremal points. In order to discern between the different cases, we equate the derivative of equation (31) , i.e. the condition for existence of the extremum:

$$\frac{27k^2}{16}Z^2-3kZ+\lambda^2+1=0 \quad or \quad Z_{1,2}=\frac{8\pm4\sqrt{1-3\lambda^2}}{9k} \tag{32}$$

The boundary value between single and triplet solutions zones corresponds to $\lambda=1/\sqrt{3}$ and referred as saddle-node (SN) bifurcation. It can be easily concluded from equation (32), SN bifurcation can take place only for positive value of parameter $k$ , for which polynomial (32) has positive roots, corresponding to real values of $N_{1,2}=\sqrt{Z_{1,2}}$ . Thus, relaxation motion exists only for hardening cubic nonlinearity, corresponds to positive value of parameter $k$ .

As a meter of fact, equation (31) defines the slow invariant manifold (SIM) of the problem. Projection of the SIM on plane $\left(N,4\left|C\left(T_1\right)\right|^2\right)$ is presented in Figure 9. The SN bifurcation points satisfy the second equation in (32). As shown in previous studies [67], [68], three-solutions zone in the SIM may allow relaxation oscillations of the slow system, represented by fast jumps from one stable branch to the other.

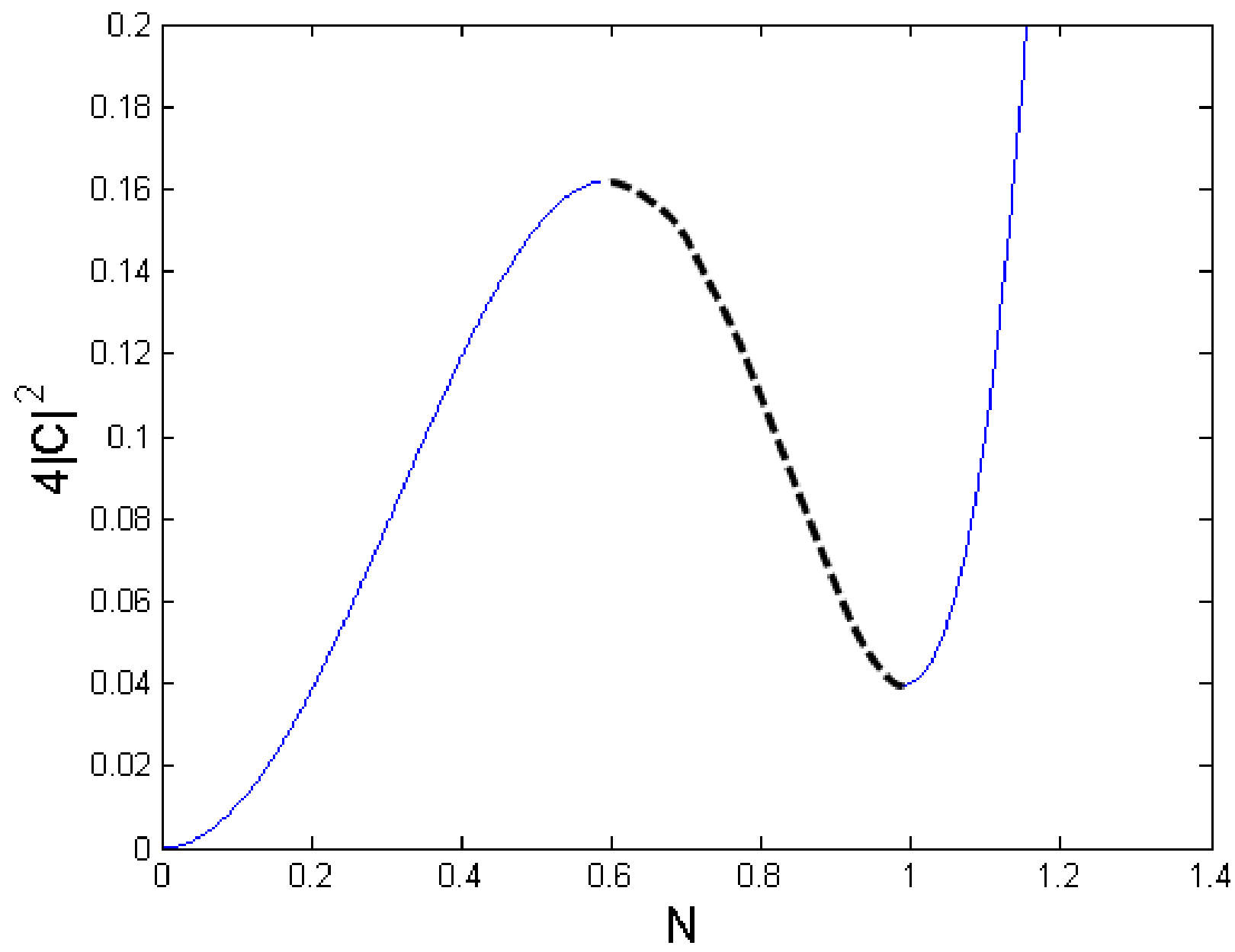

Figure 9- Projection of the SIM on plane $\left(N, 4\left|C\left(T_1\right)\right|^2\right)$ for: $k = 4/3, \lambda = 0.2$.

Description of the system motion on the SIM stable branches $\Phi\left(T_1\right) = \lim_{T_0 \to \infty} \varphi_2\left(T_0, T_1\right)$ is given by solving the $\varepsilon$ order equations of system (28) corresponding to steady-state with respect to time scale $T_0$:

$$D_1\left[\left(\frac{\lambda}{2} + \frac{i}{2}\right)\Phi - \frac{3ki}{8}|\Phi|^2\Phi\right] - \left(-\frac{1}{4}\left(1 + i\left(1-\sigma\right)\left(\lambda + i\right)\right)\Phi - \frac{3k\left(1-\sigma\right)}{16}|\Phi|^2\Phi + \frac{Ai}{4}\right) = 0 \quad (33)$$

After simple manipulations, we get the following:

$$\left(\frac{\lambda + i}{2} - \frac{3ki|\Phi|^2}{4}\right)D_1\Phi - \frac{3ki}{8}\Phi^2 D_1\bar{\Phi} = G\left(\Phi, \bar{\Phi}\right) \quad (34)$$

After taking the complex conjugate of equation (34), we get the following derivative:

$$\begin{gathered} D_1\Phi = \frac{\left(h_1 - ih_2\right)G + h_3 i\Phi^2\bar{G}}{h_1^2 + h_2^2 - h_3^2|\Phi|^4} \\ h_1 = \frac{\lambda}{2},\ h_2 = \frac{1}{2} - \frac{3k|\Phi|^2}{4},\ h_3 = \frac{3k}{8} \\ G\left(\Phi, \bar{\Phi}\right) = \frac{1}{4}\left(-\left(1 + i\left(1-\sigma\right)\left(\lambda + i\right)\right)\Phi - \frac{3k\left(1-\sigma\right)}{4}|\Phi|^2 + iA\right) \end{gathered} \quad (35)$$

By dividing variable $\Phi$ to modulus and argument by using polar transformations $\Phi\left(T_1\right)=N\left(T_1\right)e^{i\theta\left(T_1\right)}$ we get the following averaged slow flow equations:

$$
\begin{aligned}
\frac{\partial N}{\partial T_1}&=\frac{-\frac{3k}{4}A\cos\theta N^2-\lambda N+\lambda A\sin\theta+A\cos\theta}{2\left(\lambda^2+1-3kN^2+\frac{27k^2}{16}N^4\right)}\\
\frac{\partial\theta}{\partial T_1}&=\frac{-\frac{27k^2}{16}(1-\sigma)N^4+\frac{3k}{4}(1-4\sigma)N^2+\frac{9k}{4}A\sin\theta N+\left(\sigma-\lambda^2(1-\sigma)\right)+\frac{\lambda A\cos\theta-A\sin\theta}{N}}{2\left(\lambda^2+1-3kN^2+\frac{27k^2}{16}N^4\right)}
\end{aligned}
\tag{36}
$$

We denote the numerator of both equations in system (36) as $f_1\left(N,\theta\right)$ and $f_2\left(N,\theta\right)$, respectively, and their identical denominator as $g\left(N\right)$ to yield the following:

$$
\begin{aligned}
\frac{\partial N}{\partial T_1}&=\frac{f_1\left(N,\theta\right)}{g\left(N\right)}\\
\frac{\partial\theta}{\partial T_1}&=\frac{f_2\left(N,\theta\right)}{g\left(N\right)}
\end{aligned}
\tag{37}
$$

From observing system (37), one can distinguish between two different types of fixed points of the slow-flow system. The first one, refereed as regular fixed points of the slow flow, and fulfills both: $\partial N/\partial T_1=\partial\theta/\partial T_1=0$ and $g\left(N\right)\neq 0$. The points satisfying these conditions fabricate the slow invariant manifold (SIM) of the system. The second type referred to as folded singularities, and fulfills both: $\partial N/\partial T_1=\partial\theta/\partial T_1=0$ and $g\left(N\right)=0$.

Nullifying both denominators $f_i\left(N,\theta\right)$ yields the following system of equations:

$$
\begin{pmatrix}\xi_{11} & \xi_{12}\\ \xi_{21} & \xi_{22}\end{pmatrix}\begin{pmatrix}\cos\theta\\ \sin\theta\end{pmatrix}=\begin{pmatrix}\rho_1\\ \rho_2\end{pmatrix}
\tag{38}
$$

Expressions for coefficients $\xi_{ij}$ are given in Appendix E. The ordinary fixed-points are calculated from system (38) by multiplying by the inverse of matrix $\boldsymbol{\xi}$, since its determinant is $\det\left(\boldsymbol{\xi}\right)=\xi_{11}\xi_{22}-\xi_{12}\xi_{21}=-\frac{A^2}{2}g\left(N\right)\neq 0$:

$$
\alpha_3N_0^6+\alpha_2N_0^4+\alpha_1N_0^2+\alpha_4=0
\tag{39}
$$

Here $\alpha_i$ are, not surprisingly, are the same as those shown in equation(14), which describes the fixed points of the system in terms of $Z_0=N_0^2$.

The second type of fixed points, additionally to nullifying $f_i\left(N,\theta\right)$, satisfies the following condition:

$$g\left(N\right)=0\Rightarrow\frac{27k^2}{16}N_0^4-3kN_0^2+\lambda^2+1=0 \tag{40}$$

By solving the bi-quadratic polynomial (40) one obtains the following solution:

$$N_0=\sqrt{\frac{8\pm4\sqrt{1-3\lambda^2}}{9k}} \tag{41}$$

From system (38) we yield the following equations:

$$\begin{gathered}A\left(1-\frac{3k}{4}N_0^2\right)\cos\theta_0+\lambda A\sin\theta_0=\lambda N_0\\ \lambda A\cos\theta_0-A\left(1-\frac{9k}{4}N_0^2\right)\sin\theta_0=\frac{27k^2}{16}N_0^2-\frac{3k}{4}\left(1-4\sigma\right)N_0^3-\left(\sigma-\lambda^2\left(1-\sigma\right)\right)N_0\end{gathered} \tag{42}$$

Since this type of solutions corresponds to elimination of the discriminant of system (38), both equations included are linearly dependent. Thus, one can treat merely the first one, and after simple trigonometric manipulations we yield the following expressions:

$$\begin{gathered}\cos\left(\theta_0+\gamma_0\right)=\frac{\lambda N_0}{\sqrt{\left(1-\frac{3k}{4}N_0^2\right)^2+\lambda^2}}\\ \sin\gamma_0=\frac{\lambda}{\sqrt{\left(1-\frac{3k}{4}N_0^2\right)^2+\lambda^2}}\end{gathered} \tag{43}$$

Equations (43) are solvable only if their right-hand sides' absolute values do not exceed unity. Be equating to unity and using equation (41) we obtain the critical excitation amplitude required in order to allow the existence of the regime discussed:

$$A_{SMR1,2}=\lambda\sqrt{\frac{2}{|k|}}\sqrt{\frac{2\pm\sqrt{1-3\lambda^2}}{1+3\lambda^2\mp\sqrt{1-3\lambda^2}}} \tag{44}$$

Obviously, the minimal excitation amplitude for existence of SMR regime corresponds to $A_{crit,SMR}=\min\left\{A_{crit,SMR,1,2}\right\}$:

$$A_{crit,SMR} = \lambda \sqrt{\frac{2}{|k|}} \sqrt{\frac{2-\sqrt{1-3\lambda^2}}{1+3\lambda^2+\sqrt{1-3\lambda^2}}} \qquad (45)$$

Hence, in order to observe SMR regime, the following condition should hold:
$A > A_{crit,SMR}(\lambda, k)$.

## 4. Numerical results

In the following section, we compare the analytical predictions of the modulation envelope of the averaged system with numerical simulations of the full equations of motion(8). Following Figure 6(b) and

a)

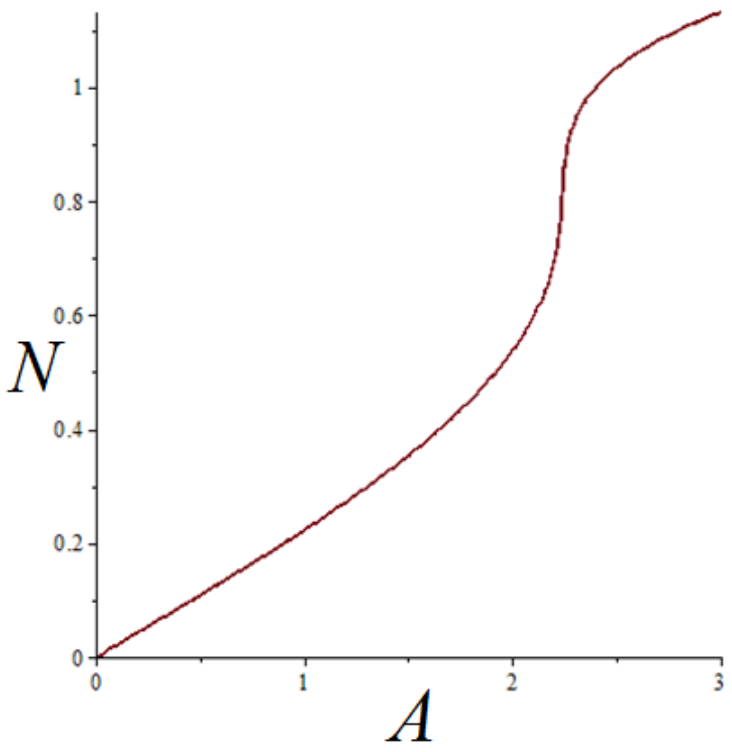

b)

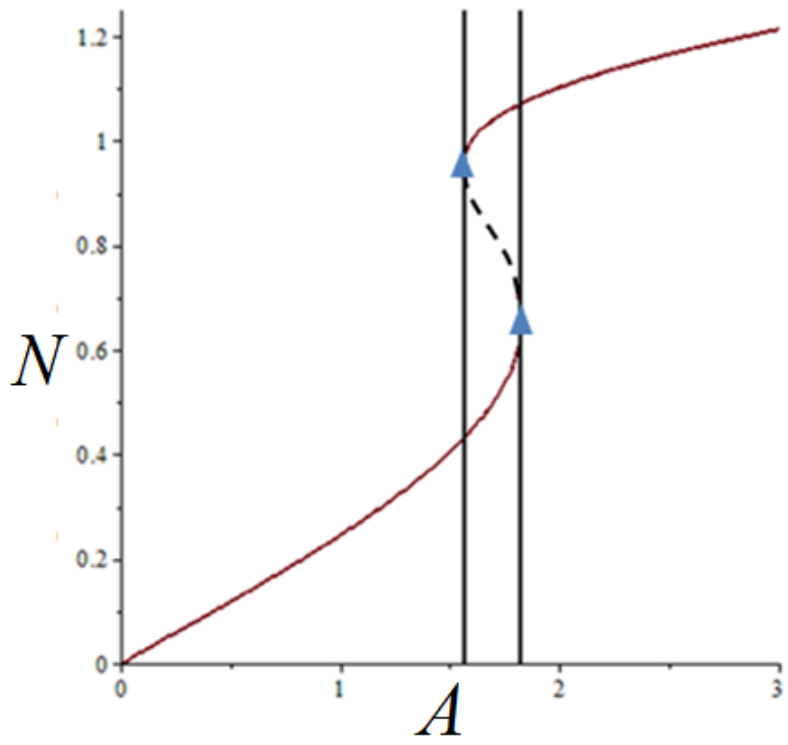

c)

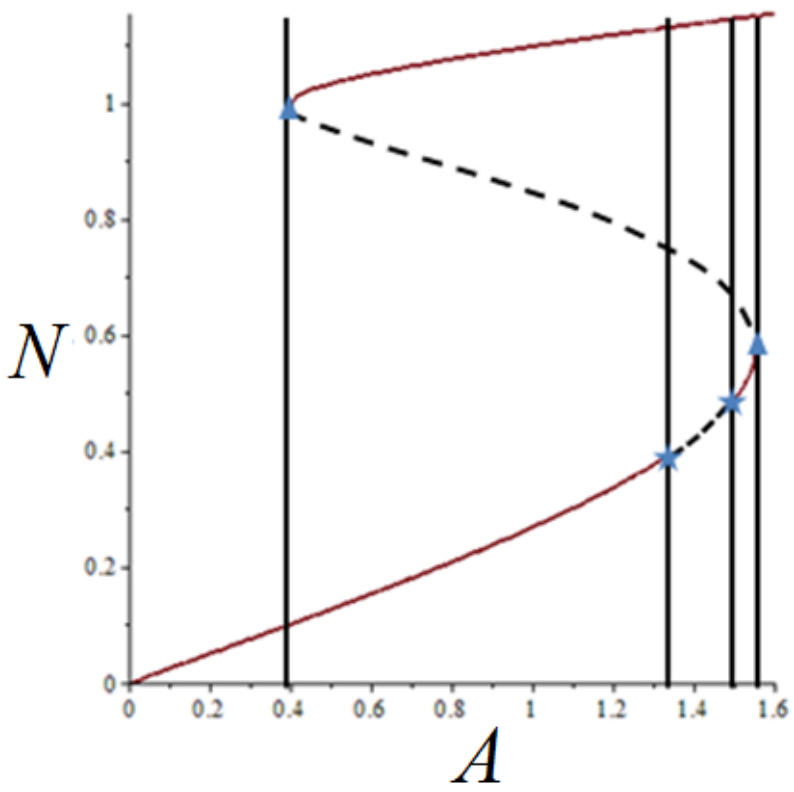

Figure 8, we examine the case corresponds the following parameter set: $k = 4/3, \sigma = 5, \varepsilon = 0.05, \beta = 5$ for different values of $A$ and $\lambda$.

The case of single periodic solution, corresponding to a)

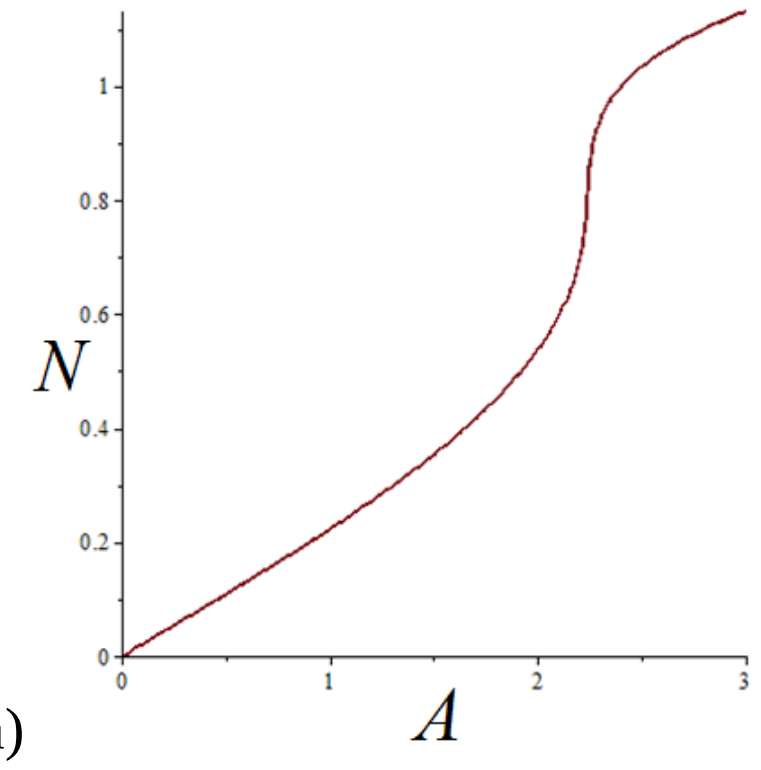

b)

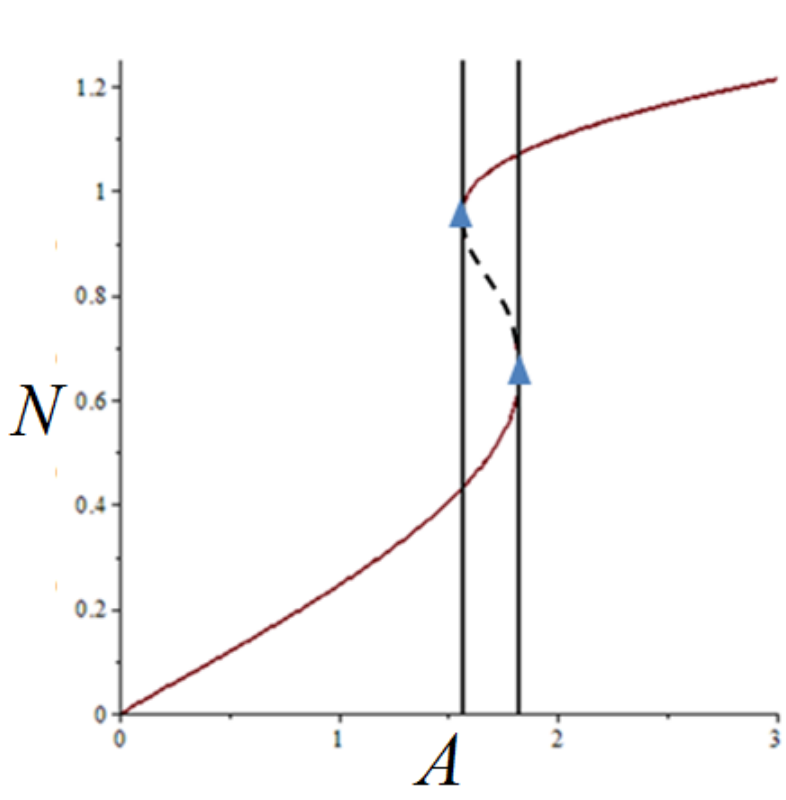

c)

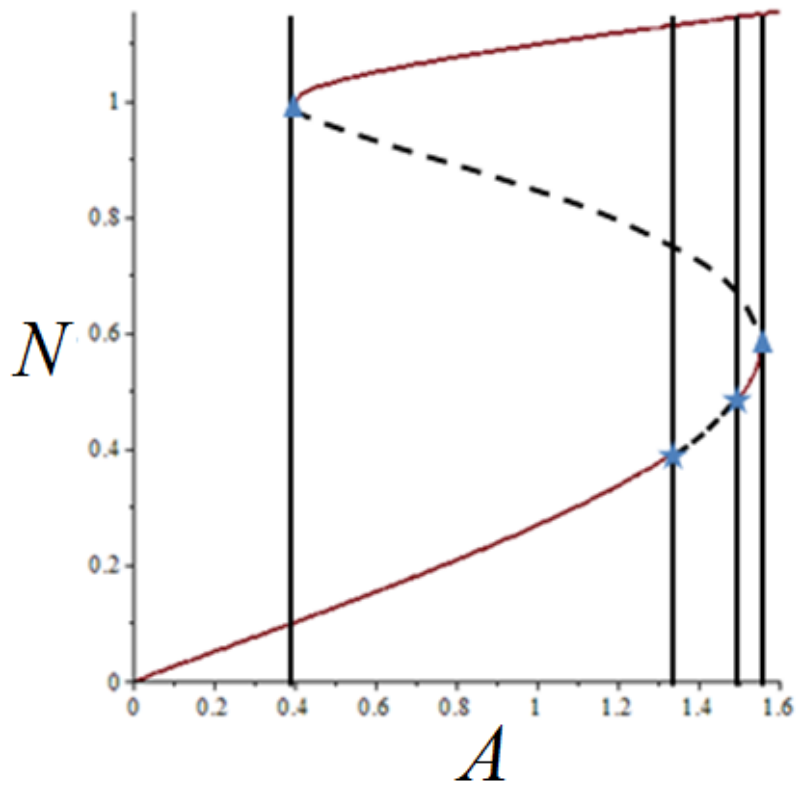

Figure 8(a), is shown in Figure 10 in terms of time history simulation and slow flow evolution on the SIM. The case of multiple periodic solution (amplitude 'jump' phenomenon), corresponding to a)

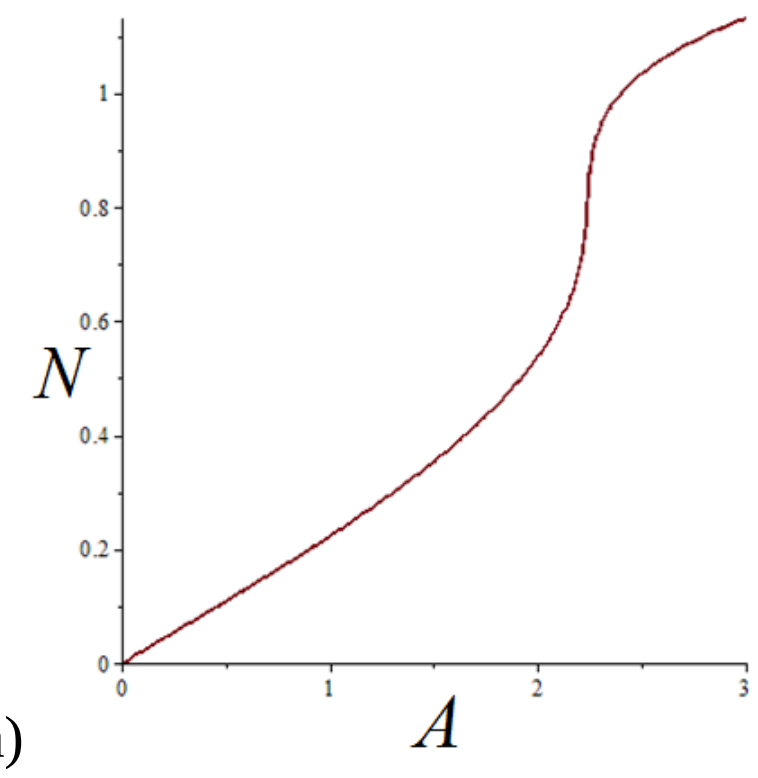

b)

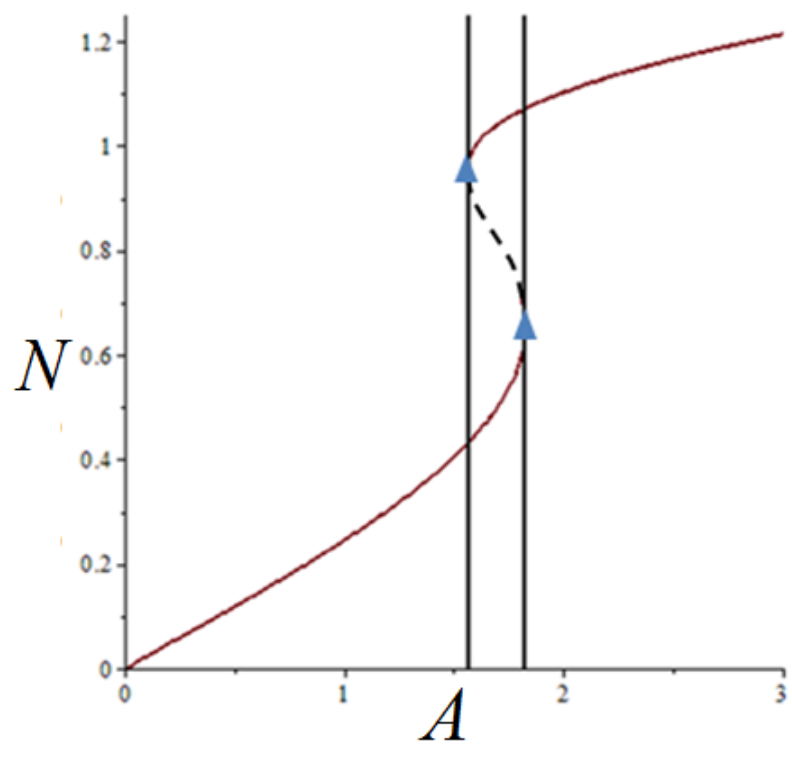

c)

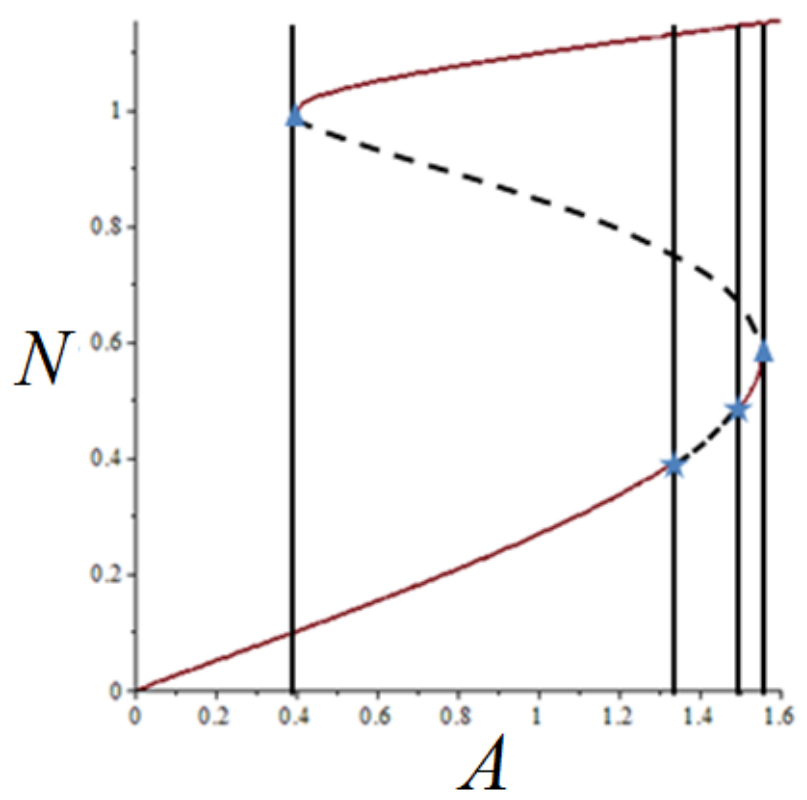

Figure 8(b), is shown in Figure 11. The case of coexistence of both periodic solution and

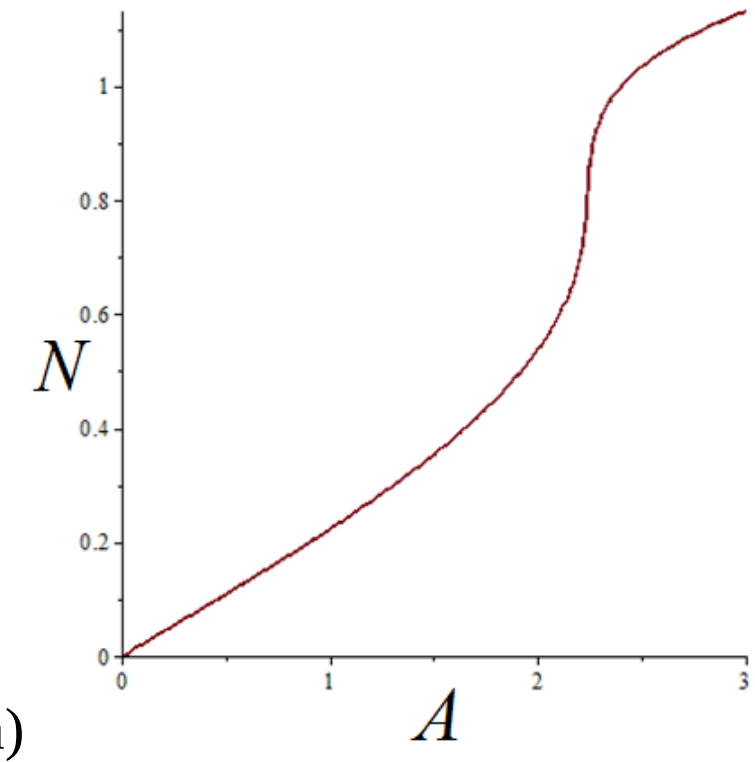

weakly quasiperiodic regime, corresponding to a) b)

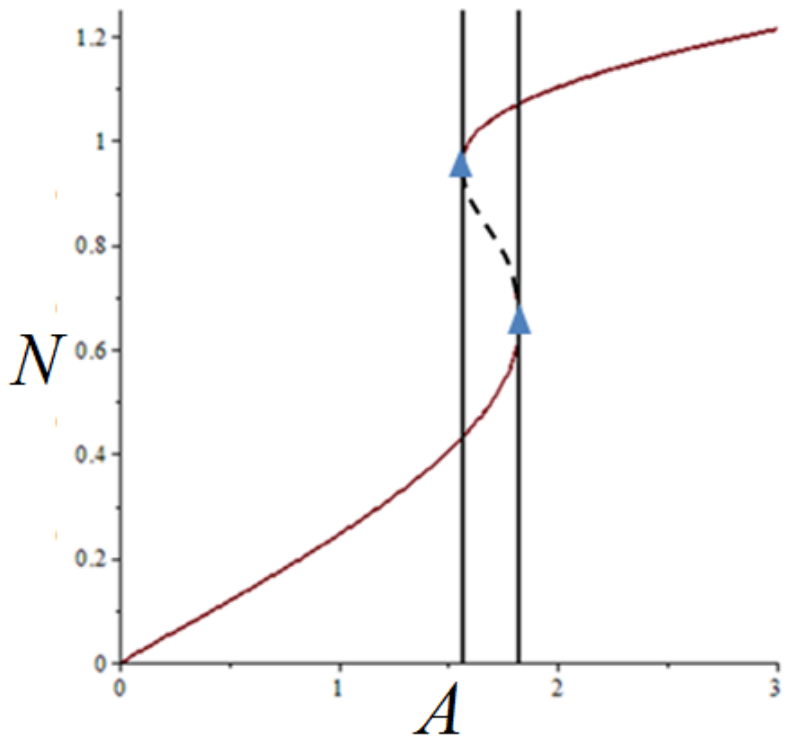

c)

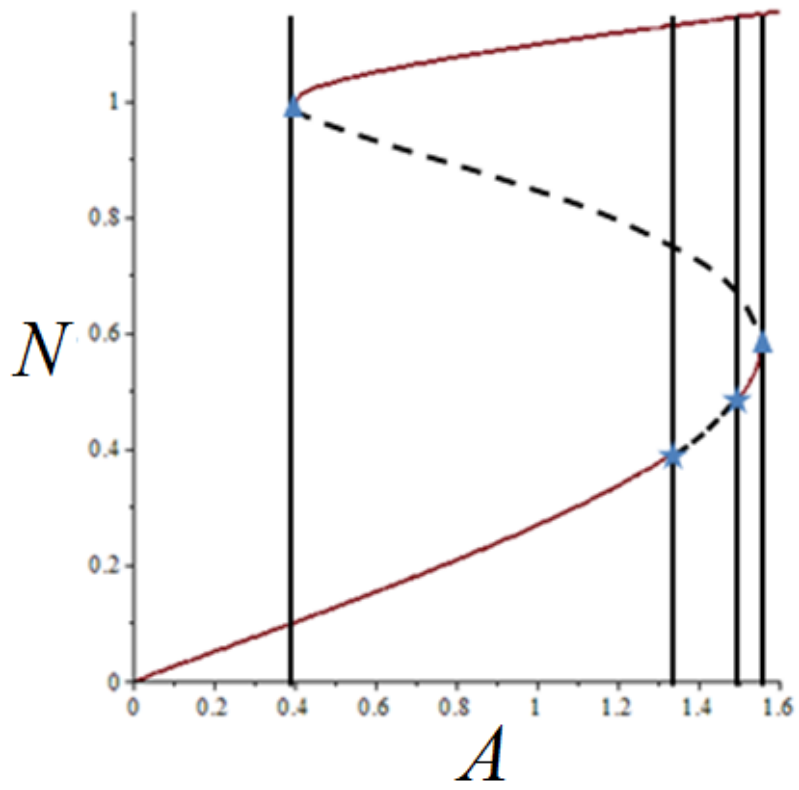

Figure 8(c), is shown in Figure 12.

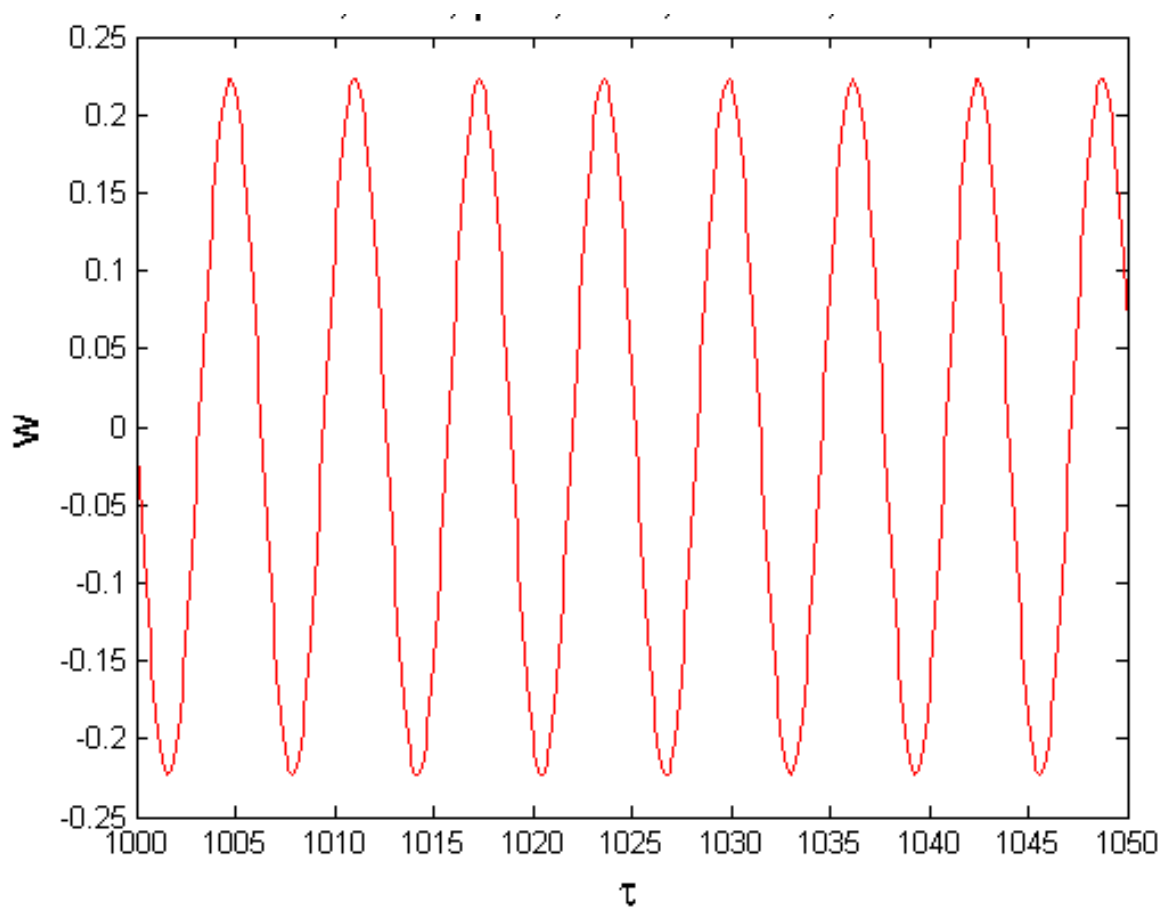

a)

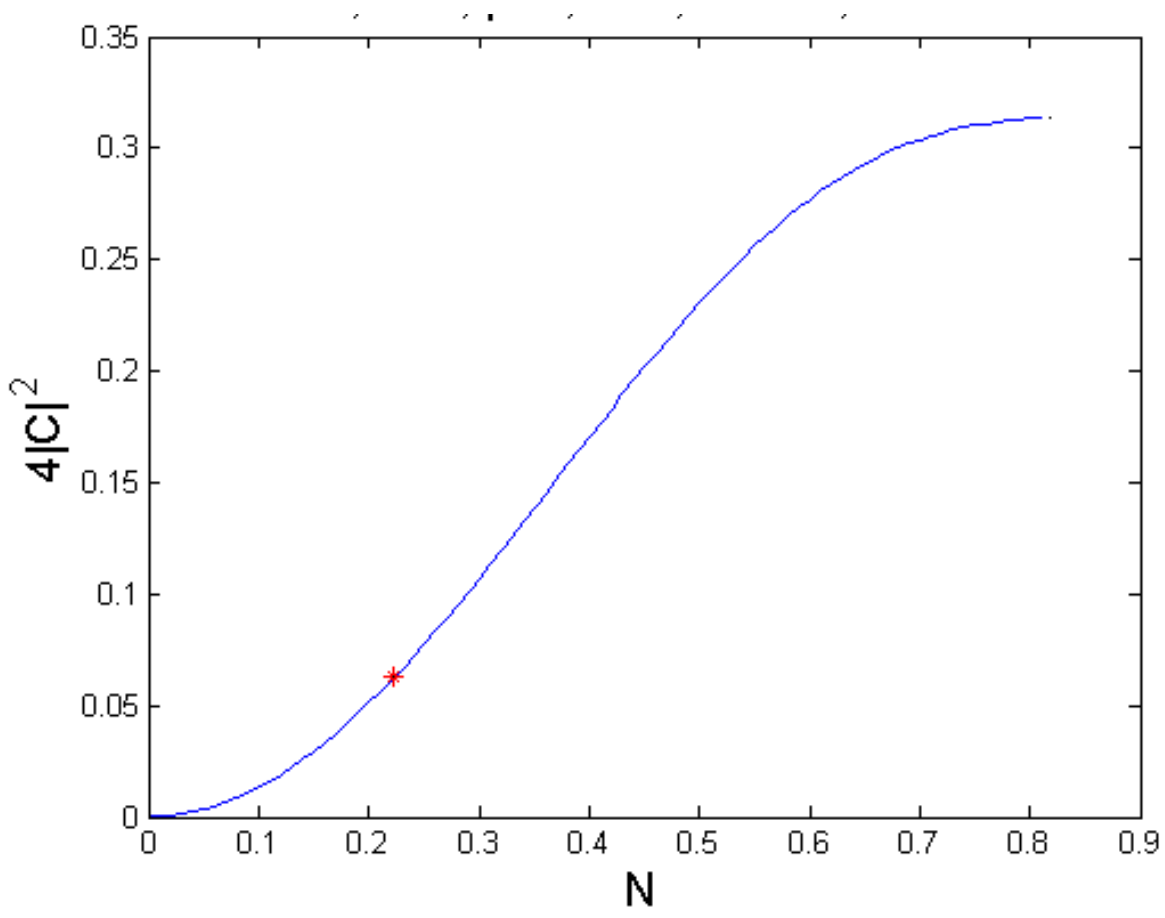

b)

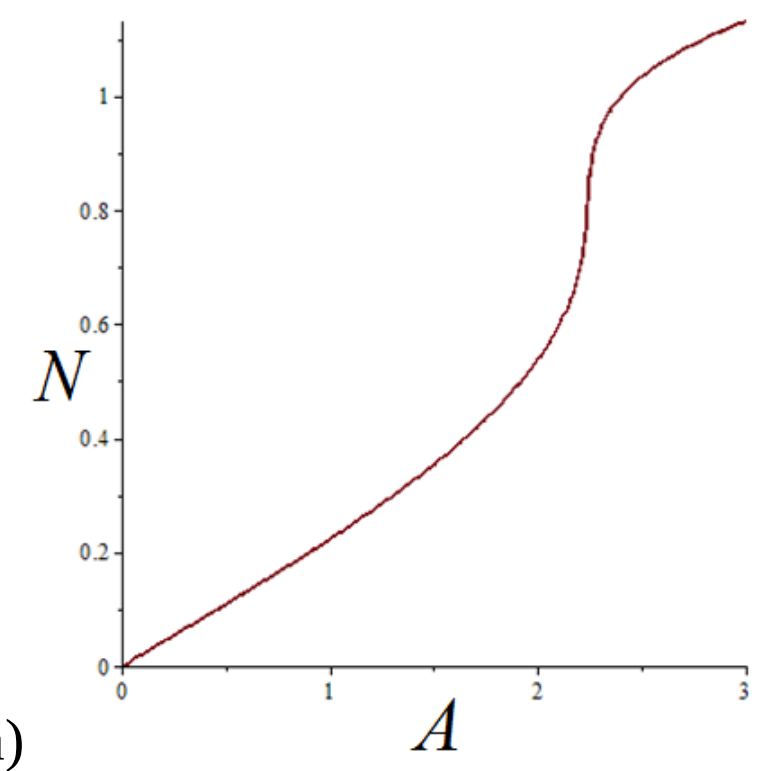

Figure 10- single periodic solution, corresponding to a) b)

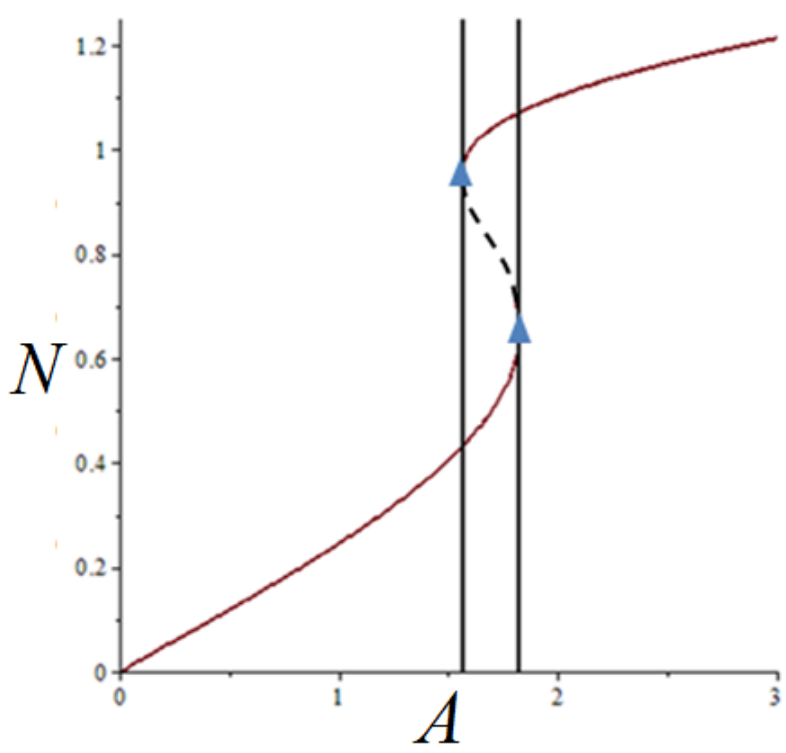

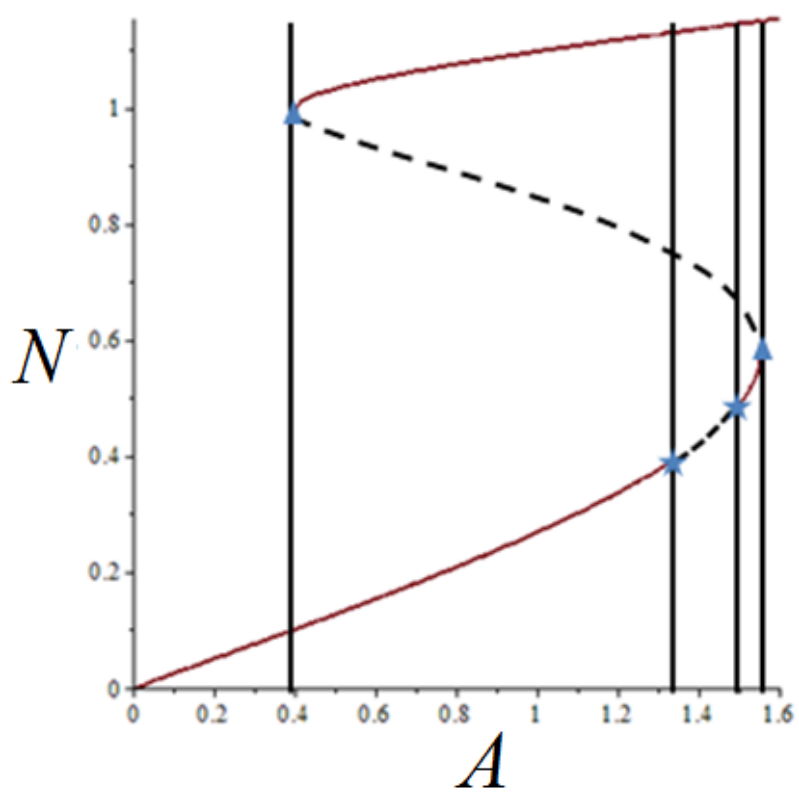

c)

Figure 8(a): $k = 4/3, \sigma = 5, \varepsilon = 0.05$, $\beta = 5 \Rightarrow \lambda_{cr,H} = 0.268,\ \lambda_{cr,SN} = 0.577$, $\lambda = 0.6 > \lambda_{cr,SN}$. Initial conditions: $u_0 = v_0 = \dot{u}_0 = \dot{v}_0 = 0$

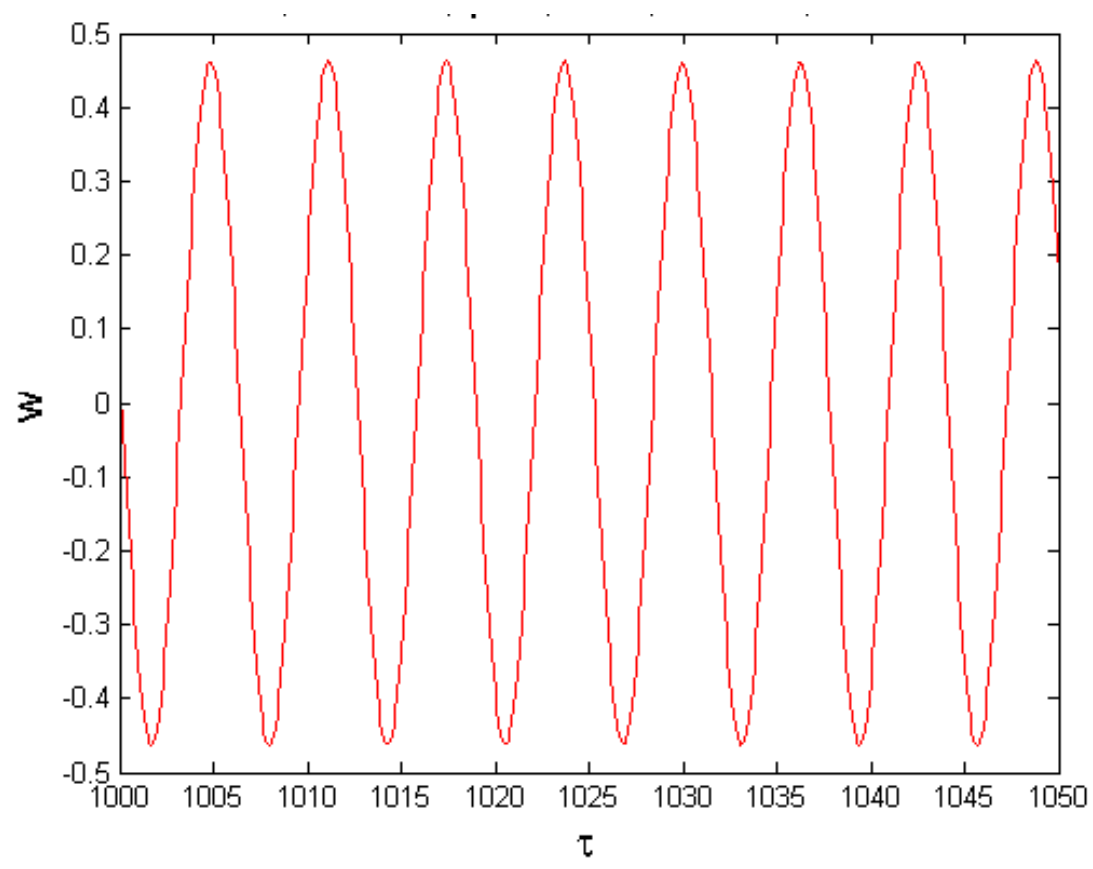

a)

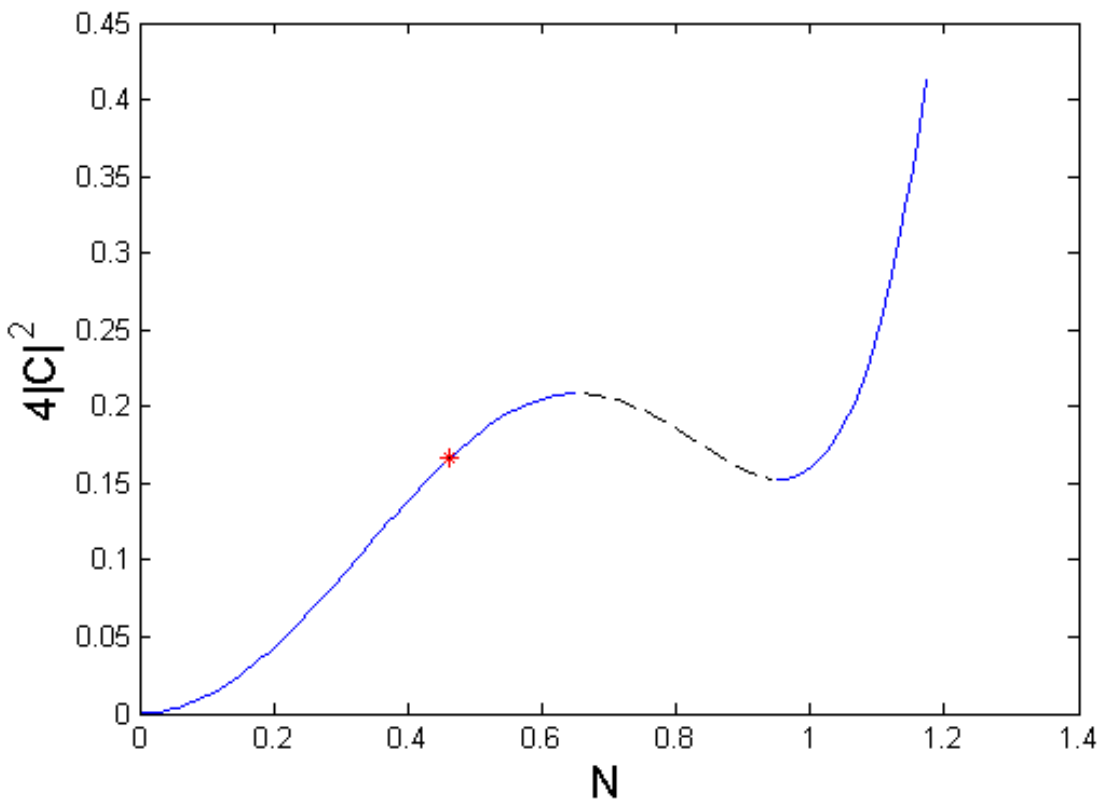

b)

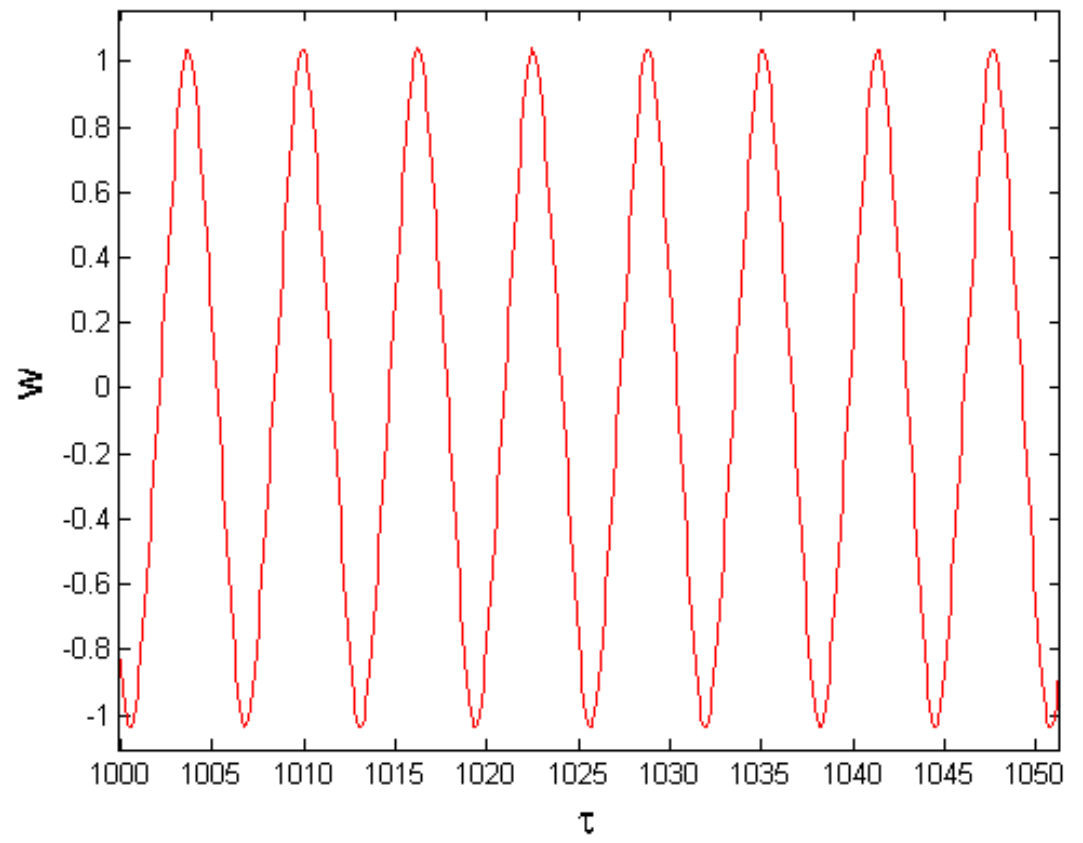

c)

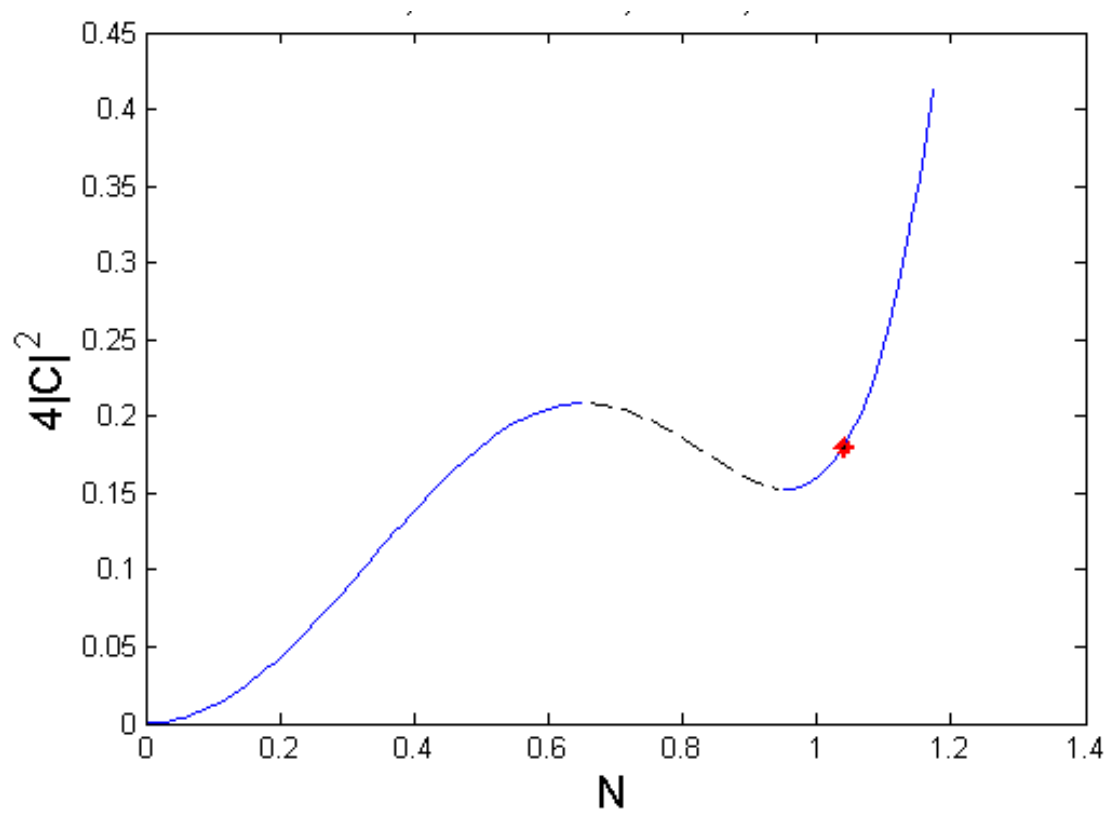

d)

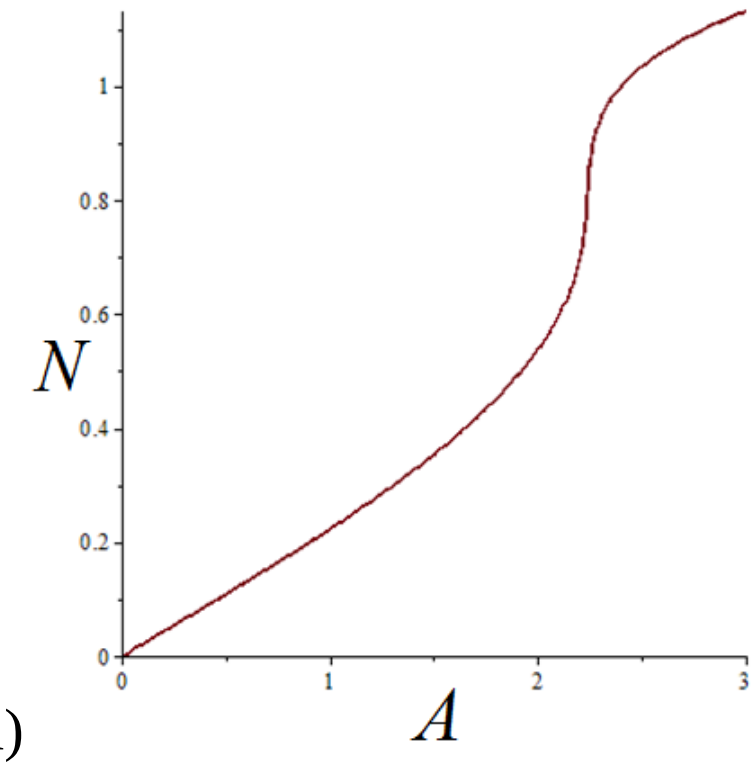

Figure 11-double periodic solution, corresponding to a) b)

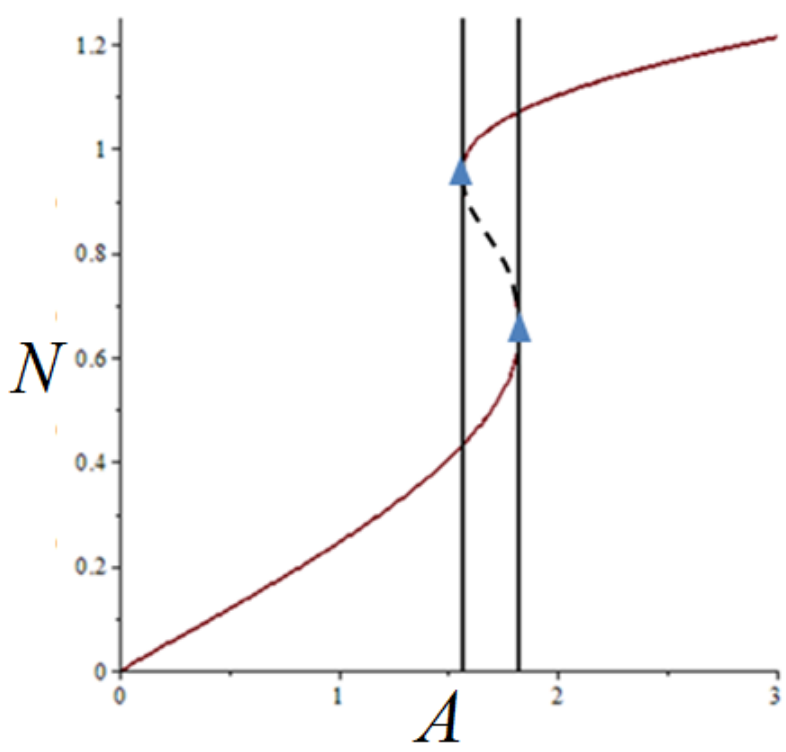

c)

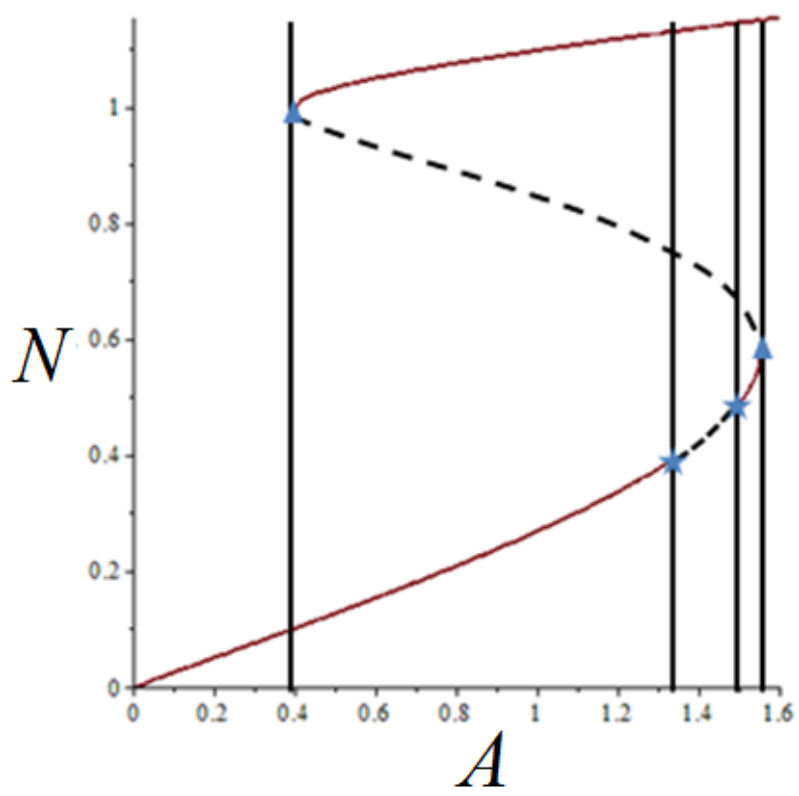

Figure 8(b): $k = 4/3, \sigma = 5, \varepsilon = 0.05$, $A = 1.62$, $\beta = 5 \Rightarrow \lambda_{cr,H} = 0.268$, $\lambda_{cr,SN} = 0.577$, $\lambda_{cr,H} < \lambda = 0.4 < \lambda_{cr,SN}$.

Initial conditions: a) $u_0 = v_0 = \dot{u}_0 = \dot{v}_0 = 0$, b) $u_0 = v_0 = 0$ $\dot{u}_0 = 0.5$ $\dot{v}_0 = 0$.

a)

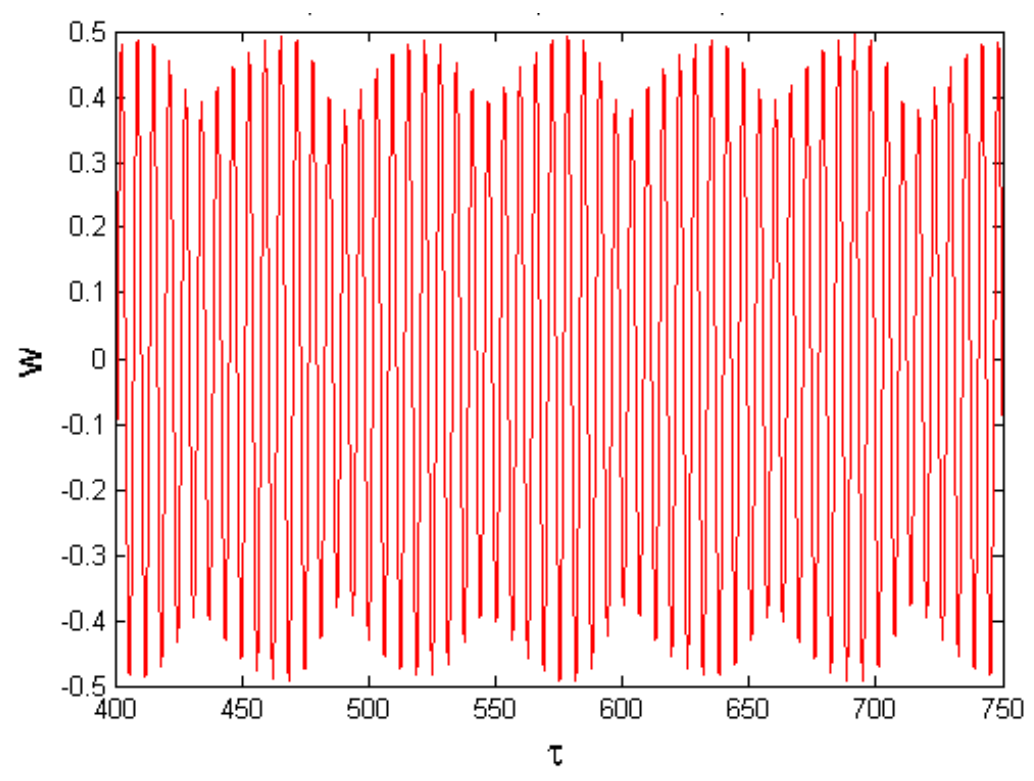

b)

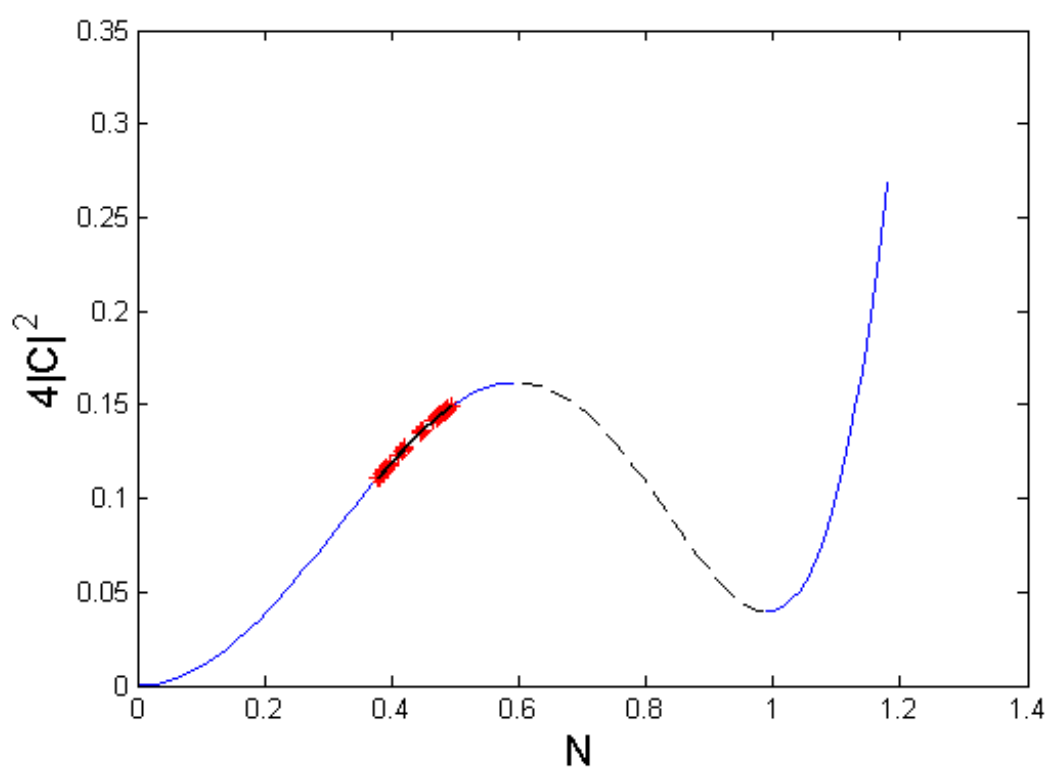

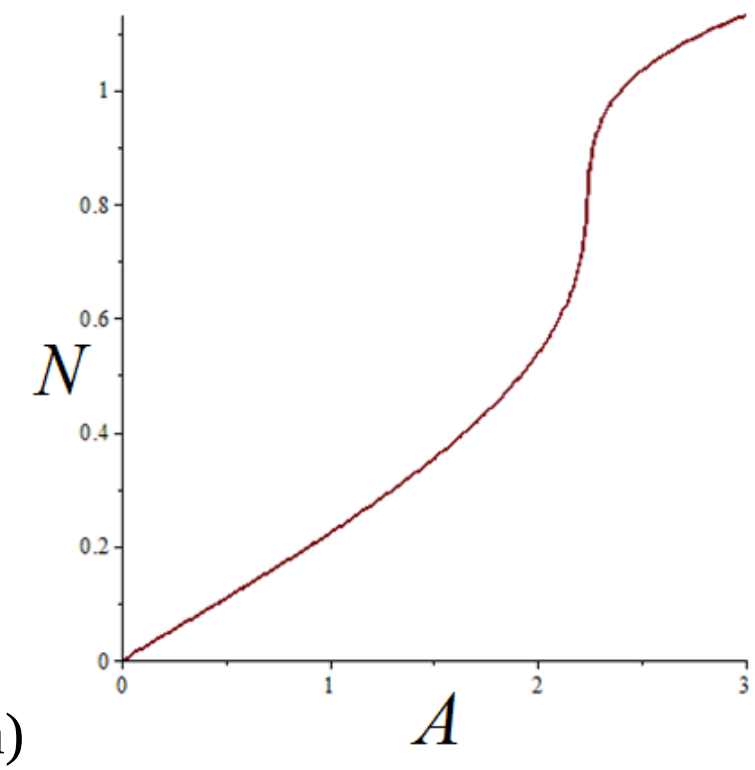

Figure 12-Quasi-periodic solution, corresponding to a) b)

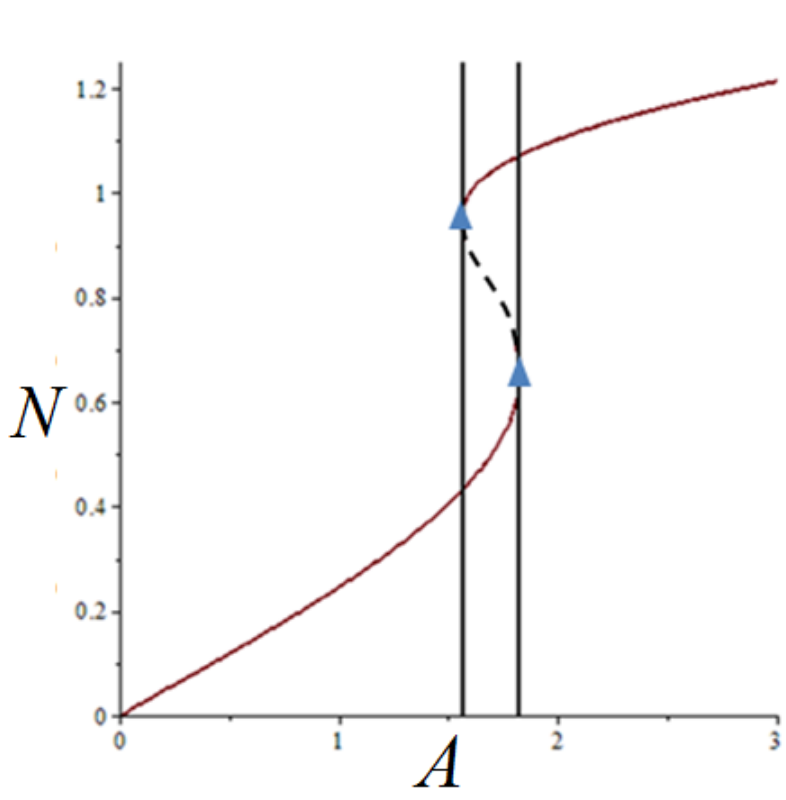

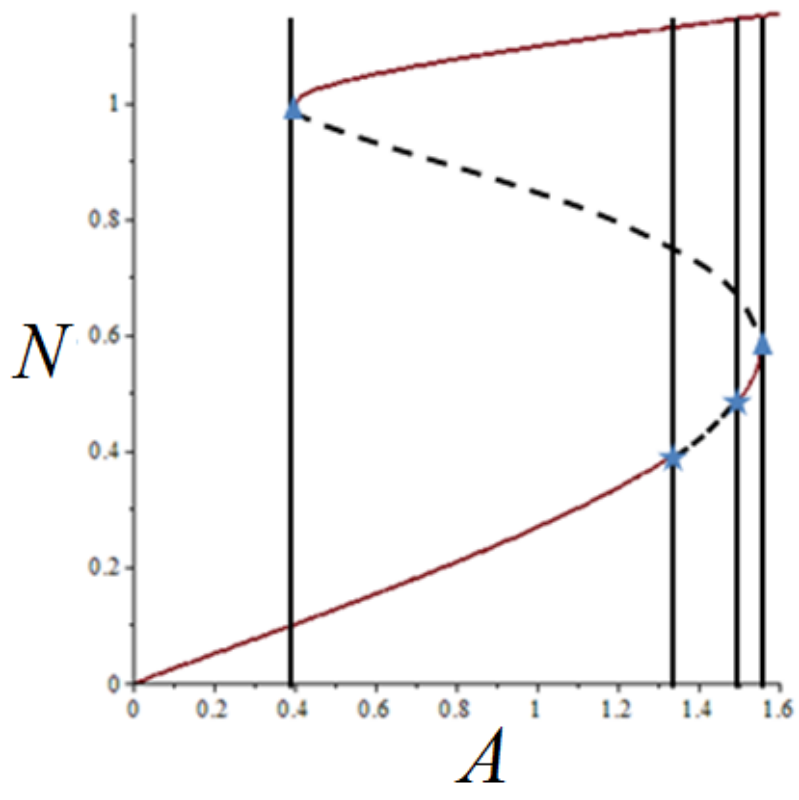

c)

Figure 8(c): $k = 4/3, \sigma = 5, \varepsilon = 0.05$, $A = 1.4$, $\beta = 5 \Rightarrow \lambda_{cr,H} = 0.268,\ \lambda_{cr,SN} = 0.577$, $\lambda = 0.1 < \lambda_{cr,H}$. Initial conditions: $u_0 = v_0 = 0, \dot{u}_0 = 0.5$ $\dot{v}_0 = 0$.

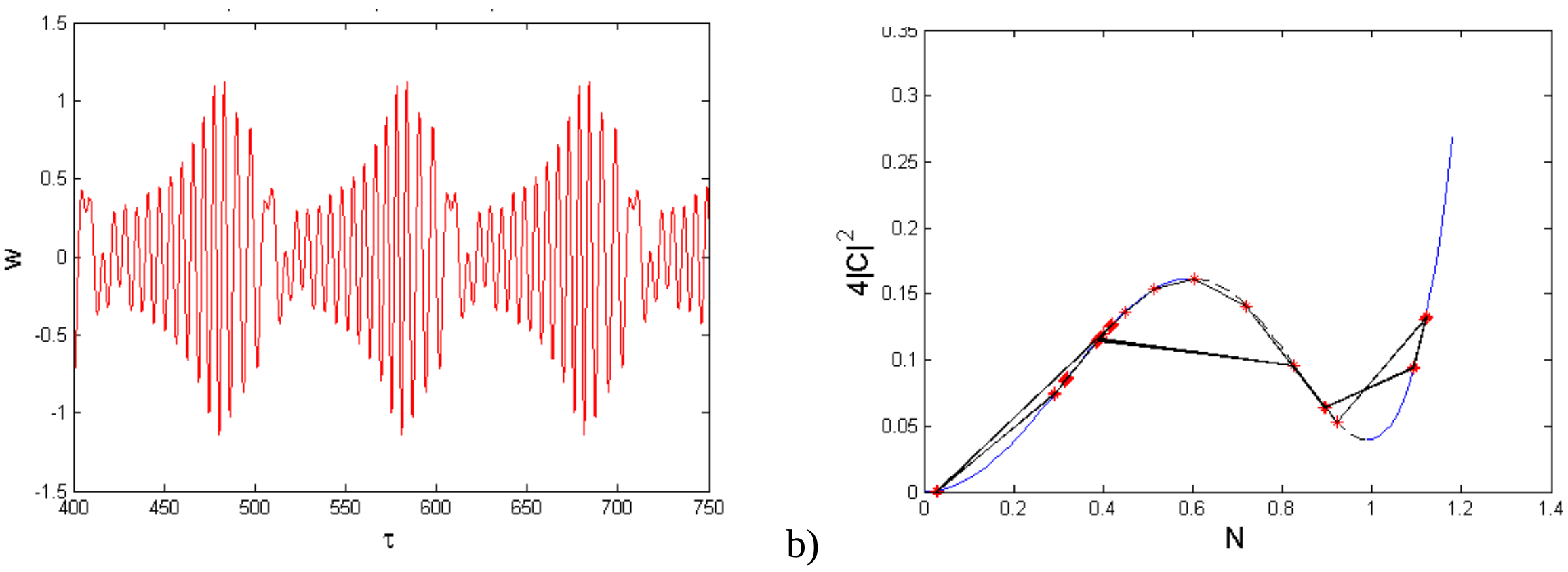

a) b)

Figure 13- SMR regime: $k=4/3, \sigma=0, \varepsilon=0.05, \beta=1$, $A=0.3$, $\lambda=0.2 \Rightarrow A > A_{cr,SMR}=0.176$. Initial conditions: $u_0=0.29$ $v_0=-0.5$, $\dot{u}_0=0.9$ $\dot{v}_0=-0.15$.

# 5. Qualitative comparison with computational and experimental results

As mentioned in the introduction, a lot of experimental and computational studies was made in the field of weakly-nonlinear response regimes analysis of liquid sloshing in partially-filled tank subjected to horizontal periodic ground excitation. The vast majority of the sloshing regimes documented are two dimensional, i.e. single periodic solutions, existence of two periodic solutions (amplitude jump), weakly-nonlinear beatings, and the SMR. In previous sections, we presented a reduced order model for partially-filled liquid vessel exposed to horizontal ground excitation in section 2. The response regimes were indicated, described and analyzed with the help of analytical and asymptotical tools in section 3. Finally, the analytical predictions were validated numerically in section 4.

In this section, the dynamical regimes revealed in previous sections are compared qualitatively with several experimental and computational studies. As one can observe in Figure 14- Figure 15 the comparison show qualitative similarity between the weakly-nonlinear dynamical regimes revealed with the help of the relatively simple reduced-order model introduced in the current study and the documented results. Amplitude 'jump' phenomenon in partially-filled liquid tanks under horizontal periodic ground motion associated with hardening and softening nonlinearities was already observed experimentally by Abramson [1] and analytically by Bauer [5].

Hill and Frandsen [14] solved analytically third-order amplitude evolution equation in terms of Jacobian elliptic functions to describe the evolution of weakly nonlinear sloshing wave, under assumptions of finite fluid depth and non-breaking waves. Classic hard-spring and soft-spring Duffing-type responses were revealed. Frandsen [3] performed fully nonlinear sloshing simulations which show good agreement for small horizontal forcing amplitude in comparison with second order small perturbation theory. For large amplitude horizontal forcing, nonlinear effects are revealed both by the third-order single modal solution and the fully non-linear numerical model. Hill [69] performed weakly-nonlinear analysis of the transient evolution of two-dimensional, standing waves in a rectangular basin using multiple-scales approach in vicinity of the critical fluid depth. The effects of detuning with respect to primary resonance, viscous damping, and cubic nonlinearity are considered. Both hardening and softening frequency response curves were produced and beating responses were indicated.

Based on multidimensional modal analysis, Faltinsen [2] investigated numerically the influence of free liquid surface initial conditions. He showed that both the calculations and the experimental results indicate that both weakly and strongly modulated nonlinear transient waves take place. The maximum liquid elevation on the tank inner wall was almost twice larger than the linear model predictions. In addition, it was shown that the nonlinear effects become more dominant with decreasing liquid level and increasing forcing amplitude due to amplification of higher modes and failure of the single-dominant modal analysis.

In the work made by Zhang et al. [4] boundary element method (BEM) adopted to investigate the sloshing in 3D rectangular tanks in vicinity of second-order resonance. Fully-nonlinear free-surface boundary conditions were applied. Fluid free-surface elevation at the left tank wall is shown in Figure 15(a). One can see that the described regime has similar nature as SMR.

Ikeda et al. [27] investigated nonlinear sloshing in rectangular tanks subjected to obliquely horizontal, harmonic excitation in vicinity of internal resonance condition 1:1 between the natural frequencies of predominate modes. Nonlinear modal equations of motion were derived, considering nine sloshing modes. Furthermore, Hopf bifurcation was documented, and amplitude modulated motions (referred to as AMMs), including chaotic motions, may appear depending on the value of the excitation frequency.

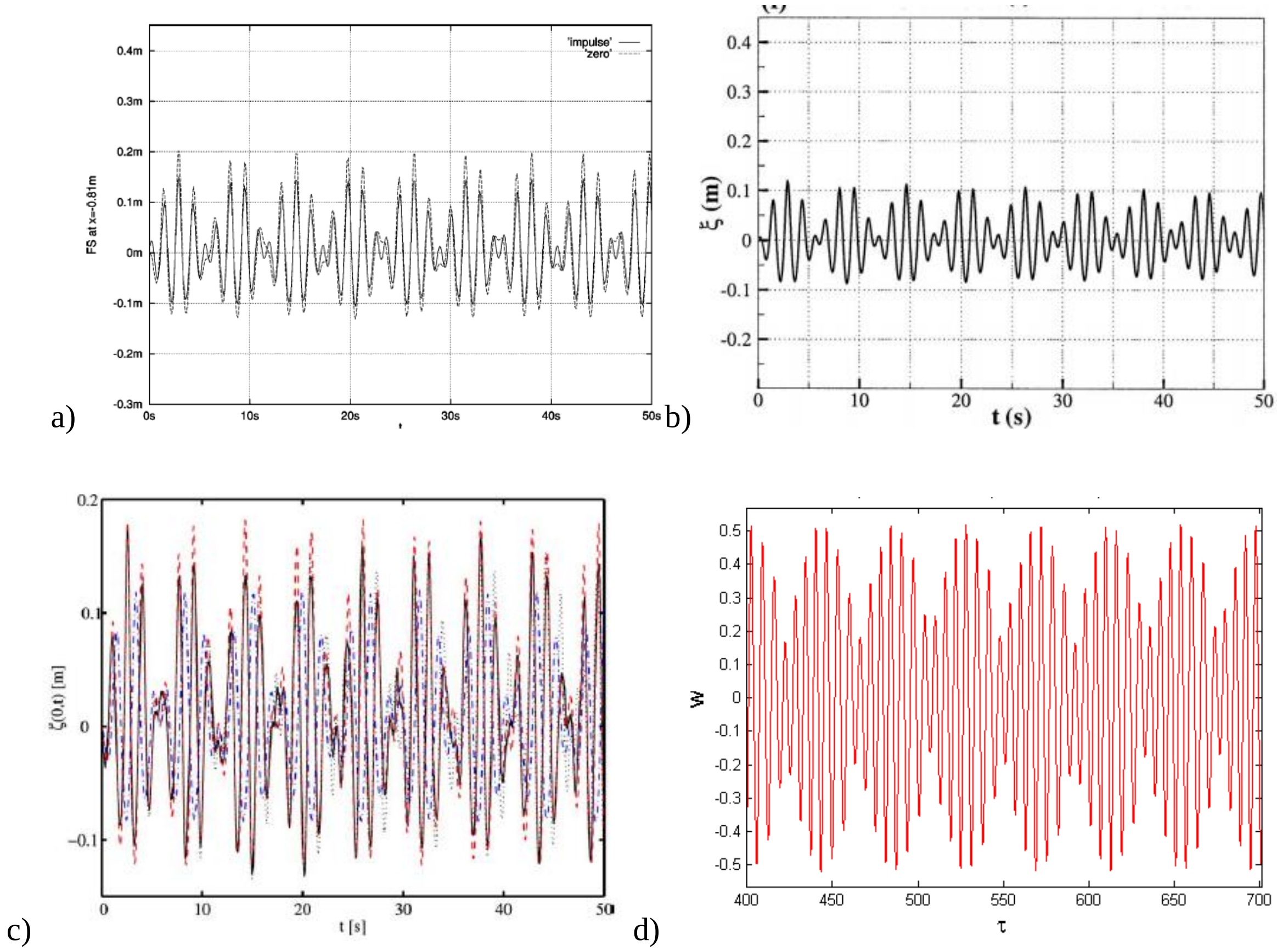

Figure 14- Nonlinear beatings response regime in partially-filled rectangular tank subjected to horizontal periodic excitation. Free-surface elevation at the left wall in horizontally excited tank ; a) Faltinsen [2] ;b) Hill [69]; c) comparison between Frandsen and Faltinsen [3]; d)current model for parameter set: $\lambda = 0.4, A = 0.3, \sigma = -0.2, \varepsilon = 0.1, k = 5$

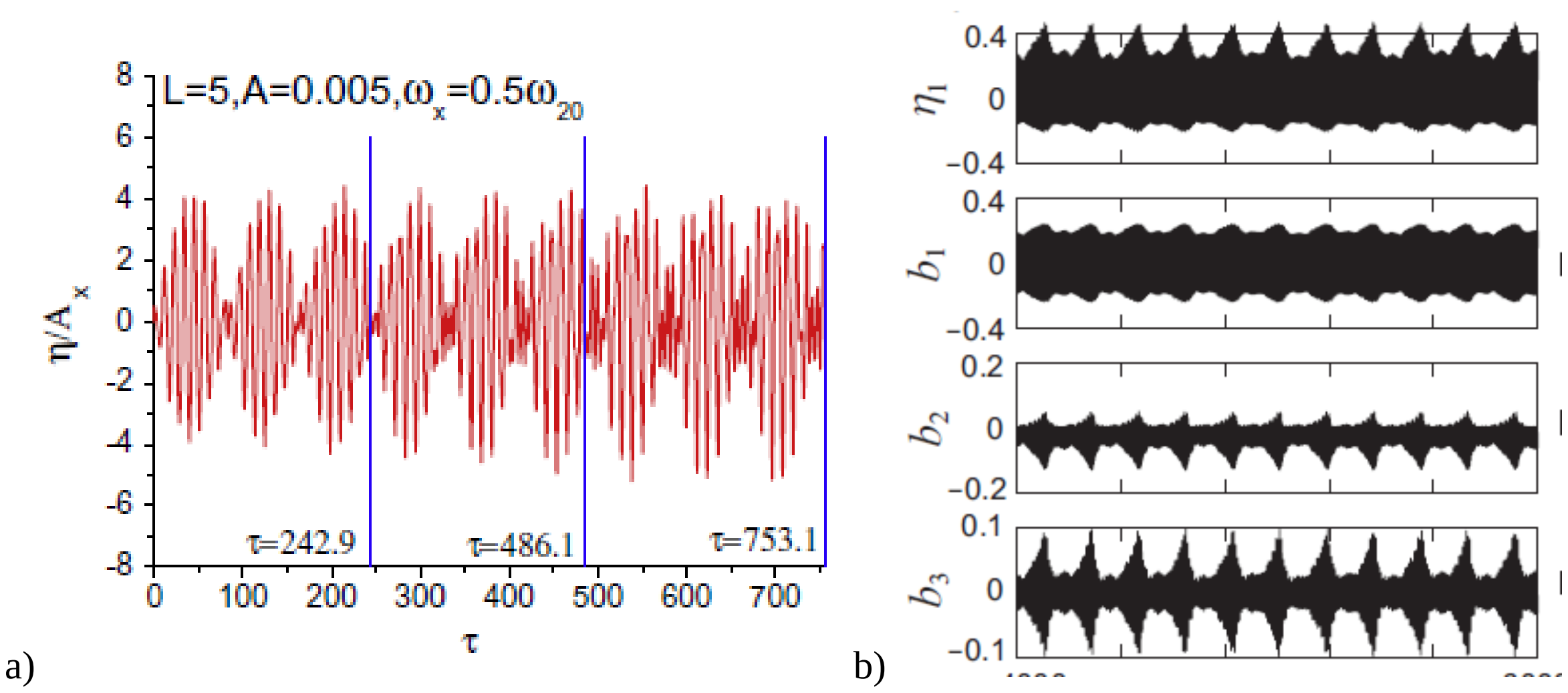

c) 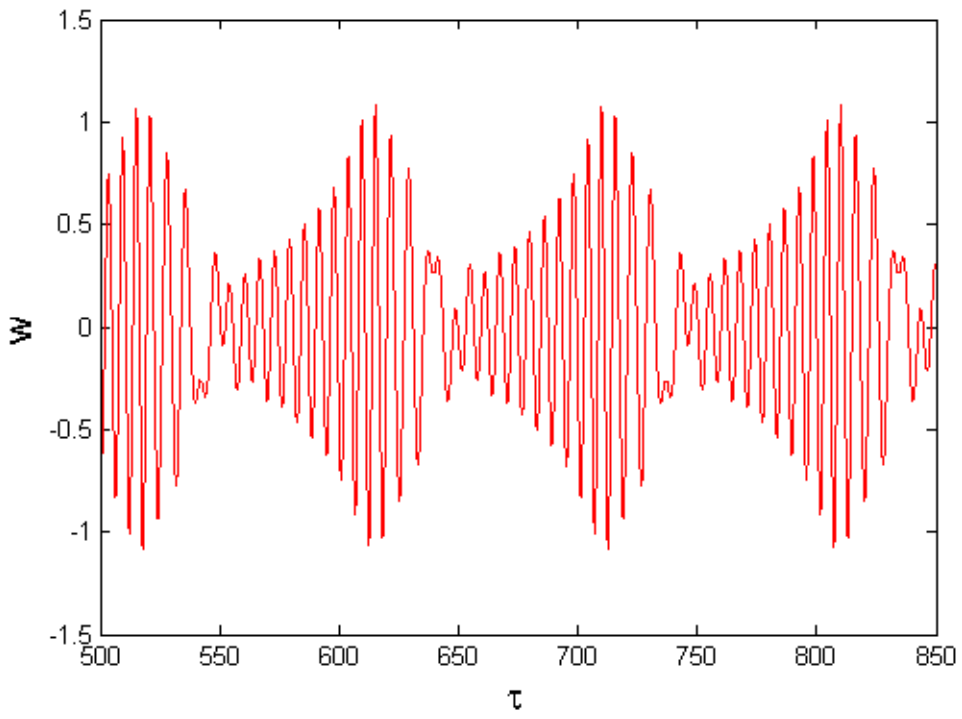

Figure 15- Strongly modulated response reigime (SMR); a) Zhang et al. [4]; b) Ikeda et al. [27]; c) Current model for parameter set: $\lambda = 0.2, A = 0.3, \beta = 1, \sigma = 0, \varepsilon = 0.05, k = 4/3$.

## 6. Concluding remarks

Novel relatively simple two dimensional design of equivalent mechanical model was suggested for qualitative description and understanding of the nonlinear two-dimensional sloshing regimes and their underlying dynamical mechanisms. It is noteworthy, that the relevance of the current model to three-dimensional sloshing regimes (e.g. swirling) was not treated herein. The basic model may be reduced to the well-studied cubic NES with addition of the linear spring; this analogy provides efficient methods for assessment of non-planar and weakly nonlinear sloshing regimes inside a partially filled vessel. The analytical predictions were validated numerically. Finally, qualitative comparison demonstrates considerable similarity between the regimes detected by this model and the regimes documented experimentally and numerically. This mechanical model paves the way towards a better stress assessment method for different engineering purposes.

## Appendix A

$$\alpha_1 = \lambda^2 + \frac{\sigma^2}{(1-\sigma)^2} + \frac{2\beta\varepsilon\sigma}{1-\sigma} + \beta^2\varepsilon^2$$

$$\alpha_2 = \frac{3k}{2(1-\sigma)}\left(\sigma + \varepsilon\beta(1-\sigma)\right)$$

$$\alpha_3 = \frac{9k^2}{16}$$

$$\alpha_4 = -\frac{A^2}{(1-\sigma)^2}$$

## Appendix B

$$c_{11} = \frac{i\varepsilon(1-\sigma)}{2(1+\varepsilon)};\; c_{12} = \frac{i\varepsilon(1+\varepsilon\sigma)}{2(1+\varepsilon)};\; c_{21} = \frac{i(1+\varepsilon\sigma)}{2(1+\varepsilon)}$$

$$c_{22} = \frac{\lambda(1+\varepsilon)}{2};\, c_{23} = \frac{i}{2}\left(\frac{(1-\varepsilon^2\sigma)}{(1+\varepsilon)} - \varepsilon\beta(1+\varepsilon)\right);\; c_{24} = \frac{3ki(1+\varepsilon)}{8}$$

## Appendix C

$$\gamma_1 = \lambda(1+\varepsilon)$$

$$\gamma_2 = \frac{27}{64}k^2(1+\varepsilon)^2 N^4 + \frac{3k(1+\varepsilon)}{4}\left(\frac{\varepsilon\sigma^2 - 1}{1+\varepsilon} + \varepsilon\beta(1+\varepsilon)\right)N^2$$
$$+\frac{1}{4}\left(\lambda^2(1+\varepsilon)^2 + \left(\varepsilon^2\sigma + 1\right) + 2\varepsilon\beta\left(\varepsilon^2\sigma - 1\right) + \varepsilon^2\beta^2(1+\varepsilon)^2\right)$$

$$\gamma_3 = \frac{1}{4}\varepsilon\lambda\left(\varepsilon\sigma^2 + 1\right)$$

$$\gamma_4 = \frac{27}{256}\varepsilon^2 k^2(1-\sigma)^2 N^4 + \frac{3k\varepsilon^2(1-\sigma)}{16}\left(\sigma + \varepsilon\beta(1-\sigma)\right)N^2$$
$$+\frac{\varepsilon^2}{16}\left(\lambda^2(1-\sigma)^2 + \sigma^2 + 2\varepsilon\beta\sigma(1-\sigma) + \varepsilon^2\beta^2(1-\sigma)^2\right)$$

## Appendix D

$$\overline{c}_1 = c_{11} + c_{22} + c_{23}$$
$$\overline{c}_2 = -c_{24}$$
$$\overline{c}_3 = c_{12}c_{21} - c_{11}\left(c_{22} + c_{23}\right)$$
$$\overline{c}_4 = c_{11}c_{24}$$
$$\overline{c}_5 = c_{11} + c_{21}$$

# Appendix E

$$\xi_{11} = A\left(1 - \frac{3k}{4}N^2\right)$$

$$\xi_{12} = \lambda A$$

$$\xi_{21} = \lambda A$$

$$\xi_{22} = -A\left(1 - \frac{9k}{4}N^2\right)$$

$$\rho_1 = \lambda N$$

$$\rho_2 = \frac{27k^2}{16}N^5 - \frac{3k}{4}(1 - 4\sigma)N^3 - \left(\sigma - \lambda^2(1 - \sigma)\right)N$$